\documentclass[twocolumn,showpacs,amsmath,prx,superscriptaddress]{revtex4-1}
\usepackage{color}
\usepackage{graphicx}
\usepackage{epsfig,psfrag}
\usepackage{amsmath}
\usepackage{amssymb}
\usepackage{amsbsy}
\usepackage{amsthm}
\usepackage{amsfonts}
\usepackage{wasysym}
\usepackage{bbm}
\usepackage{tabularx}
\usepackage{euscript}
\usepackage{enumerate}
\usepackage{amsfonts}
\usepackage{exscale}
\usepackage{bbold}
\usepackage{float}
\usepackage{slashed}
\usepackage{subfigure}
\usepackage{comment}
\usepackage[colorlinks,citecolor=blue]{hyperref}

\newcommand{\bea}{\begin{eqnarray}}
\newcommand{\eea}{\end{eqnarray}}
\newcommand{\ba}{\begin{array}}
\newcommand{\ea}{\end{array}}

\renewcommand{\Im}{{\rm Im}}

\def\bea{\begin{eqnarray}}
\def\eea{\end{eqnarray}}
\def\Tr{\mathrm{Tr}}
\def\nn{\nonumber}
\def\bea{\begin{eqnarray}}
\def\eea{\end{eqnarray}}
\def\nn{\nonumber}
\def\ba{\begin{array}}
\def\ea{\end{array}}
\def\nn{\nonumber}
\def\Tr{\text{Tr}}

\def\O{\mathcal{O}}
\def\A{\mathcal{A}}
\def\B{\mathcal{B}}
\def\C{\mathcal{C}}
\def\tri{\bigtriangleup}

\begin{document}

\title{Landau poles in condensed matter systems}

\author{Shao-Kai Jian}
\affiliation{Condensed Matter Theory Center and Joint Quantum Institute, Department of Physics, University of Maryland, College Park, Maryland 20742, USA}

\author{Edwin Barnes}
\affiliation{Department of Physics, Virginia Tech, Blacksburg, Virginia 24061, USA}

\author{Sankar Das Sarma}
\affiliation{Condensed Matter Theory Center and Joint Quantum Institute, Department of Physics, University of Maryland, College Park, Maryland 20742, USA}

\begin{abstract}
The existence or not of Landau poles is one of the oldest open questions in non-asymptotic quantum field theories. 
We investigate the Landau pole issue in two condensed matter systems whose long-wavelength physics is described by appropriate quantum field theories: the critical quantum magnet and Dirac fermions in graphene with long-range Coulomb interactions. 
The critical quantum magnet provides a classic example of a quantum phase transition, and it is well described by the $\phi^4$ theory. 
We find that the irrelevant but symmetry-allowed couplings, such as the $\phi^6$ potential, can significantly change the fate of the Landau pole in the emergent $\phi^4$ theory. 
We obtain the coupled beta functions of a $\phi^4 + \phi^6$ potential at both small and large orders. 
Already from the one-loop calculation, the Landau pole is replaced by an ultraviolet fixed point. A Lipatov analysis at large orders reveals that the inclusion of a $\phi^6$ term also has important repercussions for the high-order expansion of the beta functions. We also investigate the role of the Landau pole in a very different system: Dirac fermions in 2+1 dimensions with long-range Coulomb interactions, e.g., graphene. Both the weak-coupling perturbation theory up to two loops and a low-order large-$N$ calculation show the absence of a Landau pole. 
Furthermore, we calculate the asymptotic expansion coefficients of the beta function. We find that the asymptotic coefficient is bounded by that of a pure bosonic $\phi^4$ theory, and consequently graphene is free from Landau poles if the pure $\phi^4$ theory does not manifest a Landau pole. We briefly discuss possible experiments that could potentially probe the existence of a Landau pole in these systems. 
Studying Landau poles in suitable condensed matter systems is of considerable fundamental importance since the relevant Landau pole energy scales in particle physics, whether it is quantum electrodynamics or Higgs physics, are completely unattainable.

\end{abstract}

\maketitle

\section{Background}

Quantum electrodynamics (QED) is the most successful theory in physics. 
The most recent QED calculation up to 10th order in the fine structure coupling constant (i.e. the fifth order in the QED perturbation theory) involves the accurate determination of 389 different high-dimensional integrals contributed by 6354 Feynman vertex diagrams, with a resultant electron anomalous magnetic moment agreeing quantitatively with experiments up to 10 significant digits~\cite{Nio2018}. Obviously, we have come a long way (although it took 70 years for this progress) from Schwinger's ground-breaking analytical work of 1948 calculating just the first order QED correction, which obtained the electron anomalous magnetic moment correct to two significant digits~\cite{Schwinger1948}. 
This amazing agreement between theory and experiment is the most impressive success of quantum mechanics, and the common belief is that this astounding success will continue up to many orders in the QED perturbation theory, with theory and experiment agreeing certainly up to well over 100 significant digits, although neither experiment nor theory is likely to get to such a high precision in the foreseeable future. 
Because the QED perturbation theory is asymptotic, it will eventually break down, with the perturbation series eventually diverging at some very high order ($\gg 137$), but this is not a serious concern at this stage. 
The infinite-order perturbative QED result is thus bound to be incorrect since it gives a divergent answer.

In addition to the asymptotic nature of the QED perturbation theory, there is a second fundamental `problem' with QED, first emphasized by Landau~\cite{Landau1954}, and often referred to as the Landau pole (it also is sometimes called `Landau ghost' or `Moscow zero')~\cite{LandauPole}. 
The term Landau pole refers to the divergence of a running (or renormalized) coupling constant at a finite energy in a field theory. 
Landau poles happen only in field theories that are not asymptotically free, where the running coupling increases with increasing energy, in contrast to asymptotically free field theories where the coupling constant decreases with increasing energy. If the Landau pole is a true feature of QED, then the only way to obtain a reasonable theory would be to set the bare charge to zero, resulting in a theory that is completely trivial (or noninteracting). 
Landau poles have been discussed extensively in the context of both QED and scalar field theories, such as the $\phi^4$ theory, which is relevant for Higgs bosons in the standard model. 
The problem is, however, that the existence of a Landau pole is theoretically established only within the leading-order perturbative approximation (extended to very high energies) and therefore, whether the Landau pole is real or an artifact of the perturbation theory is unknown. 
In the context of QED, direct numerical simulations have been used to explore the existence of Landau poles, but whether the poles have any physical relevance in the sense of imposing triviality remains unclear~\cite{Stuben1998, Gies2004, Djukanovic2018}. 
Unfortunately, approaches based on lattice simulations are hindered by chiral symmetry breaking, making the parameter regime where a Landau pole occurs inaccessible~\cite{Stuben1998}. Numerical methods have also been applied to $\phi^4$ theory, with the conclusion that the continuum theory is most likely trivial. This finding has in turn been used to impose bounds on the Higgs mass~\cite{Dashen1983, Weisz}. However, the triviality question has not been fully resolved here either, and a stronger bound on the mass comes from unitarity~\cite{Lee1977}, obscuring the role of the Landau pole. It is also unclear to what extent these findings carry over to effective scalar theories, where higher-order couplings allowed by symmetry can affect renormalization group (RG) flows at high energies. 
It is possible that a fully nonperturbative, strong-coupling theory would be necessary to eventually settle the question since weak-coupling perturbation theories simply may not be applicable in predicting the physics of a divergent renormalized coupling. 
The Landau pole problem is intrinsically tied to the asymptotic behavior of the QED perturbation expansion and to the nonperturbative effects of instantons. Although there has been recent progress in developing a more rigorous treatment of nonperturbative effects like instantons using resurgent trans-series \cite{Dunne2015},  
the question of the existence or not of the Landau pole in QED remains as open today as it was in the early 1950s when Landau first proposed it.

One thing is, however, clear. 
Even if the Landau pole exists in QED and/or quartic scalar field theories, there is no hope for its experimental manifestation in particle physics because the energy scale for the Landau pole is unphysically huge (e.g. way above the Planck scale in QED and the Higgs mass in the $\phi^4$ theory). 
Quite apart from the fact that such large energy scales are experimentally unattainable (not only now, perhaps ever), new physics, outside the scope of QED, comes in at high energy scales, and the predictions of QED for the Landau pole become academic since QED itself (quite apart from a perturbation theoretic analysis of QED leading to the Landau pole) is no longer a correct description of nature at such high energies.

This is the context of the current theoretical work, where we investigate the condensed matter analogs of the Landau pole in the theories of critical quantum magnets and the physics of graphene. 
It is well known that the $\phi^4$ scalar field theory describes the long-wavelength critical behavior of quantum magnets, and the two-dimensional massless Dirac theory describes the long-wavelength behavior of graphene. 
Thus, graphene is an example (with suitable modifications) of QED, whereas quantum magnets are examples (with suitable modifications) of the $\phi^4$ scalar field theories. 
Our goal is to study the presence/absence of Landau poles in these two concrete condensed matter physics examples to motivate further theoretical work that could shed light on the fundamental issue of triviality in the quantum field theories which are not asymptotically free. 
Another equally important objective of our work is to motivate experimental work in condensed matter systems to directly probe the existence or not of Landau poles in these two systems, which are described by continuum field theories containing Landau poles in the leading-order perturbative analysis. 
The question of the existence or not of Landau poles is of sufficient fundamental significance that anything we can learn from condensed matter systems about the possible presence/absence of Landau poles would be valuable for future progress in the subject.

\section{Introduction}
The concept of Landau poles is a long-standing issue in quantum field theory that raises questions about fundamental aspects of the renormalization group (RG), in particular the asymptotic behavior of renormalized couplings~\cite{Landau1954, Landau1955, Shirkov1976}. 
In a quantum field theory that is not asymptotically free, consider the beta function that governs the running of a dimensionless coupling constant $\lambda$,
\bea
	\frac{d \lambda}{d \log \mu} = \beta(\lambda),
\eea
where $\mu$ is the energy scale of interest. 
One can integrate over both sides to get
\bea
	\int_{\lambda(\mu_1)}^{\infty} \frac{d \lambda}{\beta(\lambda)} = \int_{\mu_1}^{\mu_\infty} d\log \mu.
\eea
$\mu_\infty$ is the energy scale at which the running coupling becomes infinite. 
Suppose the beta function has the asymptotic behavior $\beta(\lambda) \propto  \lambda^a, \lambda \gg 1$,
where $a$ is a constant independent of $\lambda$, then
\bea
	\int_{\lambda(\mu_1)}^{\infty} \frac{d \lambda}{\beta(\lambda)} \propto \int_{\lambda(\mu_1)}^\infty \lambda^{-a} d\lambda = 
	\begin{cases} \infty, \quad & a \le 1 \\
			      \log\frac{\mu_\infty}{\mu_1} < \infty, \quad & a > 1.
	\end{cases} \nn\\\label{eq:Landaupoletest}
\eea
In the first case, $a \le 1$, there is no Landau pole. 
The running coupling reaches infinity only at infinite energy scales. 
In the second case, $a>1$, however, the coupling diverges at a finite energy scale $\mu_\infty$. This is a Landau pole. 
Clearly, the existence of a Landau pole depends on the asymptotic form of the beta function at $\lambda\gg 1$.

Because Landau poles were first discovered theoretically in quantum field theories relevant for high energy physics, such as QED, research into this issue has remained largely within the high-energy physics community~\cite{Callaway1988,Callaway1986,Djukanovic2018,Kim2002,Gies2004,Stuben1998,Lizzi2013}. 
The simplest solution to the Landau pole problem is the idea of quantum triviality~\cite{Callaway1988, Callaway1986, Kim2002, Gies2004}, in which the pole is avoided by setting the coupling to zero at all scales, yielding a trivial, noninteracting theory. This is of course unsatisfactory if one is interested in the effect of interactions (which are clearly present in QED experiments), so a more phenomenological resolution that is often adopted is to argue that the theory is incomplete and gets replaced by another theory at high energies before the Landau pole is reached. 
For instance, QED is usually believed to be just one part of a more fundamental electroweak theory. 
Moreover, the Landau pole in QED can be estimated to occur at an energy scale on the order of $10^{286}$ eV, which is far beyond the Planck scale $10^{28}$ eV, suggesting that it is a purely academic issue. However, the reliability of such estimates is questionable given that the Landau pole is intrinsically tied to the large-coupling regime, whereas most analyses rely on weak-coupling perturbation theory carried out to first or second order. Many attempts have been made to go beyond small-order perturbation theory starting from Lipatov's method~\cite{Lipatov1977, Zinn-Justin1981} for calculating large orders in an asymptotic series. In fact, a resummation procedure based on such large-order expansions suggests that the Landau pole does not exist at all in the $\phi^4$ theory~\cite{Suslov2001, Suslov2008}. 
But the Landau pole question is by no means settled since the resummation technique is essentially a Borel interpolation, and the exact strong coupling theory remains elusive, so one cannot be sure that the Landau pole does not exist. 
In particular, numerical simulations seem to indicate its existence~\cite{Gies2004}.

In addition to having wide application in particle physics, continuum field theories and the renormalization group are also important for condensed matter systems, especially in the context of phase transitions, where the universal physics is controlled by long-wavelength fluctuations at a critical point~\cite{Kadanoff1966, Wilson1975}. 
(In fact, Wilson developed his RG theory motivated by condensed matter considerations in critical phenomena, and the first problem he solved using his momentum-shell RG theory is the Kondo problem, a celebrated condensed matter problem involving singular spin-flip scattering of electrons in metals from quenched magnetic impurities~\cite{Wilson1971}.) 
Thus, it is natural to consider the Landau pole problem in condensed matter systems. 
Unlike most cases in high-energy physics, here the ultraviolet completion is well understood as it comes from the lattice, which brings a natural energy scale cutoff set by the inverse lattice constant. The question then becomes whether a Landau pole arises above or below this scale. This question is particularly important in systems where the coupling constant can be large, as is the case in graphene, where the fine structure constant away from the Dirac point is $\alpha \sim 1$~\cite{Geim2009}, and even more so in synthetic twisted bilayer graphene, where $\alpha \sim 10$~\cite{MacDonald2011, Cao2018a, Cao2018b, SDS2019}. The interactions in these systems are substantially stronger than in QED, where $\alpha \approx 1/137$, suggesting that if there is a Landau pole, then its energy scale could occur below the cutoff, potentially leading to consequences that are experimentally accessible. 
We emphasize that in condensed matter physics, all continuum field theories are, by definition, effective field theories since the lattice explicitly breaks the continuum by providing a short-distance cutoff length scale (or equivalently, an effective ultraviolet energy/momentum cutoff for the system which does not have to be put in by hand in the theory).  The field theory is valid up to this ultraviolet energy/momentum scale ($\sim$the inverse lattice constant), which is a physical constraint defining the domain of validity of the effective field theory. 
Our current understanding of all field theories as effective field theories up to some scale is explicitly obeyed in condensed matter physics and does not have to be artificially introduced as, for example, in lattice gauge theories. 
Of course, one does not know the precise cutoff scale except that it should be set dimensionally by the lattice spacing. 
It is possible that the cutoff is in fact somewhat shorter (longer) than the lattice spacing, in which case the effective field theory would apply to energies above (below) the inverse lattice spacing. 
In the context of Landau poles, the hope is that the pole energy here would not be much higher than the inverse lattice spacing energy (within an order of magnitude) so that there is some reason to believe that the effective field theory does not completely break down at the Landau pole energy. 
In such a situation, it is sensible to consider condensed matter experiments to investigate the presence/absence of Landau poles. 
If the pole happens to be at an energy much larger than the natural ultraviolet cutoff for the lattice in the system, there is no hope for experimentally studying the Landau pole in the corresponding condensed matter system since the effective field theory is unlikely to apply at that high-energy scale. 

Here, we consider two condensed matter systems: the critical quantum magnet and Dirac fermions in 2+1 dimensions with long-range Coulomb interactions (graphene). Both systems are believed to be well described by effective continuum field theories. The critical quantum magnet is described by a bosonic $\phi^4$ theory as dictated by symmetry. In four dimensions, the $\phi^4$ theory has a coupling that grows logarithmically with energy scale, eventually leading to a Landau pole as determined from perturbation theory~\cite{Scammell2015}. However, this theory is an effective field theory, meaning that all terms allowed by symmetry, e.g., all even-order terms $\phi^{n}$, where $n \in 2\mathbb{Z}$, are in principle present in the action. Although the $\phi^4$ term is the most relevant one deep in the infrared, the $n>4$ terms can become important at the higher energy scales where the Landau pole potentially arises. 
This motivates us to investigate to what extent Landau poles are affected by infrared-irrelevant terms in the $\phi^4$ effective field theory. To answer this question, we study the beta function of a $\phi^4 + \phi^6$ potential and find that the Landau pole can be strongly affected by the presence of the $\phi^6$ potential at both small and large orders of the asymptotic expansion. 

Dirac fermions in graphene constitute a second important case study because the Fermi velocity approaches zero in the ultraviolet, enhancing the role of Coulomb interactions at high energies and potentially inducing a Landau pole~\cite{Geim2009}. 
The theory is infrared-stable because the Fermi velocity increases as the energy scale approaches zero, leading to a decreasing running coupling (which is cut off at some exponentially low energy scale when retardation effects due to the finite speed of light come into consideration). 
The marginal Fermi liquid behavior that results from the logarithmic growth of the Fermi velocity near the Dirac point is well studied~\cite{Vozmediano1994, Vozmediano2011, Elias2011}. 
We review the result for the small-order beta function in graphene~\cite{SDS2014, Son2007} and show that the Landau pole problem is not resolved by either weak-coupling or large-$N$ perturbation theories at one loop. 
On the other hand, a two-loop analysis using either ordinary perturbation theory or a large-$N$ expansion indicates the absence of a Landau pole. 
We also compute the large-order expansion coefficients nonperturbatively, showing that the asymptotic coefficients are smaller than those of the pure $\phi^4$ theory. This indicates that if the pure $\phi^4$ theory does not manifest a Landau pole, then neither does the Coulomb-interacting graphene field theory. 
Our main, but tentative,  conclusion in this work is that in all likelihood, Landau poles do not exist in condensed matter systems at any energy scale, although the leading-order perturbative RG theory may imply their existence, often at energy scales beyond which the corresponding effective field theory is applicable.

The paper is organized as follows. We discuss Landau poles in the critical quantum magnet and in graphene in Secs.~\ref{CQM} and~\ref{DFCI}, respectively. In each section, we compute both the small- and large-order terms of the beta function for the running coupling. While the small-order beta function is obtained through the standard perturbation theory, the large orders are evaluated through the saddle-point approximation following the Lipatov method~\cite{Lipatov1977, Zinn-Justin1981}. 
In Sec.~\ref{DD}, we briefly discuss possible experiments and conclude the paper. 
The appendixes contain technical details and an instructive zero-dimensional toy model to illustrate the steps of the Lipatov method for the $\phi^4 + \phi^6$ potential.

\section{Critical quantum magnets \label{CQM}}

In this section, we compute the small- and large-order terms of the asymptotic expansion of the beta function for the effective theory consisting of both $\phi^4$ and $\phi^6$ terms. We use the Wilsonian renormalization group to obtain the one-loop beta function. We show that while the pure $\phi^4$ theory exhibits a Landau pole, the inclusion of the $\phi^6$ term removes the pole and replaces it with an ultraviolet fixed point. In addition, we compute the large-order expansion of the beta function by using a saddle-point approximation in the regime of negative couplings. We find that the expansion coefficients of the $\phi^4$ coupling $\lambda_4$ grow as $k!$ while those of $\lambda_6$ grow as $(k!)^2$. From both small-order and large-order expansion coefficients, we see that including the $\phi^6$ term strongly affects the fate of the Landau pole compared to a pure $\phi^4$ theory.

\subsection{Small orders}
We consider the effective theory,
\bea\label{action}
	S[g_4,g_6; \phi] = \int d^d x \left[ \frac12 (\partial \phi)^2 + \frac{g_4}{4} \phi^4 + \frac{g_6}6 \phi^6 \right],
\eea
where $\phi_i$, $i=1,...,N$ is an $O(N)$ field, and the summation over $i$ is implicit: $\phi^2 = \sum_i^N (\phi_i)^2$, and $(\partial \phi)^2 = \sum_i^N \sum_\mu^d (\partial_\mu \phi_i)^2 $. 
Using the standard Wilsonian RG technique~(see Appendix~\ref{app:RG}), the perturbative beta equations at the critical surface $\bar r = 0$, where $\bar r$ is the dimensionless mass, are
\bea
\label{RG2a}	\frac{d \lambda_4}{d\log \mu} &=& -(3N+12) \lambda_6 + (4N+32) \lambda_4^2, \\
\label{RG2b}	\frac{d \lambda_6}{d\log \mu} &=& 2 \lambda_6 + (12N+168) \lambda_4 \lambda_6 - \frac{32 N + 832}3 \lambda_4^3,
\eea
where $\lambda_4 = \frac{A_{3}}{(2\pi)^4} \frac{g_4}4$ and $\lambda_6 = \Big(\frac{A_{3}}{(2\pi)^4}\Big)^2 \frac{g_6}6 \Lambda^2$ are dimensionless couplings, and we have set $d=4$. $A_d$ is the area of a $d$-sphere with unit radius.

\begin{figure}
	\includegraphics[width=5cm]{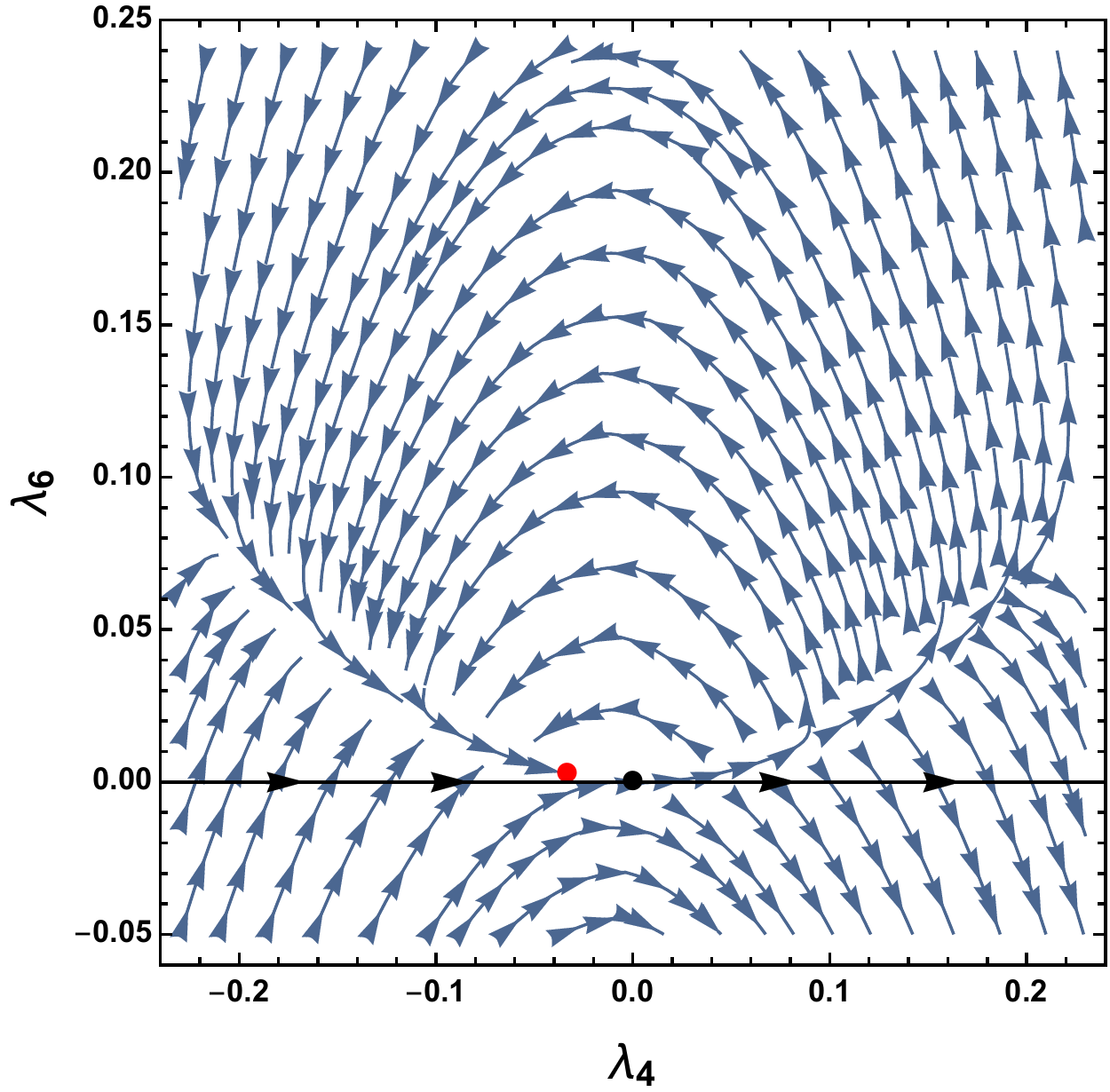}
	\caption{\label{flow1}The flow diagram of $\lambda_4$-$\lambda_6$. The blue arrow indicates the RG flow to higher energies. The black and red dots are the Gaussian and the new fixed point in Eq.~(\ref{FP1}), respectively. The dashed line which separates different flow regimes is given by Eq.~(\ref{dashed}). We take $N=1$ for simplicity. }
\end{figure}

From Eqs.~(\ref{RG2a}) and~(\ref{RG2b}), besides the Gaussian fixed point, there is also an ultraviolet fixed point at 
\bea\label{FP1}
    && (\lambda_4,\lambda_6)\nn\\
	&& =\left(- \frac{N+8}{2(N^2+6 N + 128)}, \frac{(N+8)^3}{3(N+4)(N^2+ 6 N + 128)^2} \right) \nn \\
	&& \approx \left( - \frac1{2N} + O(1/N^2), \frac1{3N^2} + O(1/N^3) \right).
\eea
The fixed point has two relevant directions at the critical surface. 
One should distinguish it from the tricritical point. 
For $d\ge3$ spacetime dimensions, the tricritical point is the same as the Gaussian fixed point, because the upper critical dimension is three for the $\phi^6$ theory. 
Moreover, a tricritical point has only one relevant direction at the critical surface.

The flow diagram of $\lambda_4$-$\lambda_6$ is shown in Fig.~\ref{flow1}, where the black and red dots are the Gaussian fixed point and the ultraviolet fixed point in Eq.~(\ref{FP1}), respectively. 
The dashed curve is given by
\bea\label{dashed}
	f(\lambda_4)= \frac{16 (N+26)\lambda_4^3}{3(6 N \lambda_4  + 84 \lambda_4 + 1)}, \quad \lambda_4>0.
\eea
If one fixes $\lambda_6 =0$ to reduce to the pure $\phi^4$ theory, as indicated by the black line in Fig.~\ref{flow1}, $\lambda_4$ will become infinite at the Landau pole. This is consistent with the one-loop beta function for $\lambda_4$. 
However, in the global flow diagram of both $\lambda_4$ and $\lambda_6$, the points at $\lambda_6 = 0$ are not stable at higher energies. 
Depending on the initial values, the RG flow regimes are separated by the dashed line given by Eq.~(\ref{dashed}). 
Denote the initial values by $\lambda_{40}$ and $\lambda_{60}$. 
On the one hand, at $\lambda_{40} <0 \cup [\lambda_{40}>0 \cap \lambda_{60} > f(\lambda_{40})]$, it will flow to the new fixed point (if one can fine tune $\lambda_i =0$ for $i=8,10,...$). 
On the other hand, at $\lambda_{40}>0\cap \lambda_{60}< f(\lambda_{40}) $, $\lambda_{6}$ will flow to negative values, indicating that the theory is not stable because the energy is not bounded from below. 
Therefore, if one includes the symmetry allowed $\phi^6$ term, the Landau pole is replaced by a fixed point of the one-loop beta function. 
Note that even if the $\phi^6$ term is not included initially, it will be generated by the $\phi^4$ potential.

\subsection{Large orders}

We have shown that the inclusion of the $\phi^6$ term in the effective theory will replace the Landau pole by an ultraviolet fixed point even in the one-loop calculation. 
We now extend the analysis to large orders. 
Namely, we calculate the asymptotic expansion coefficients $\beta_k, k\gg 1$ of beta functions, $\beta(\lambda) = \sum_k \beta_k \lambda^k $. 
We first calculate the asymptotic expansion of the correlation function, and then relate it to the beta function. 
In Appendix~\ref{app:0D}, we present a toy model that provides a simple illustration of the basic steps we employ.

\subsubsection{Asymptotic expansions of the correlation function}

Consider the $n$-point correlation function of the massless $\phi^4+\phi^6$ theory defined by Eq.~(\ref{action}),
\bea
	G_n(g_4,g_6;X_n) = \int D\phi \prod_{j=1}^{n}\phi(x_j) e^{-S},
\eea
where $X_n=(x_1, ..., x_n)$ denotes all $n$ spacetime arguments, and in this section we consider $N=1$ for simplicity. 
We are interested in computing the asymptotic expansion of the $n$-point correlation function $G_n(g_4, g_6)$ (in the following, we suppress the argument $X_n$ for notational simplicity). 
In general, the expansion has the following form:
\bea
	G_n(g_4, g_6) = \sum_{k \ge 0} A_k^{(n)} g_4^k + \sum_{k' \ge 1} B_{k'}^{(n)}(g_4) g_6^{k'}.
\eea

In four dimensions, the asymptotic expansion of $G_n(g_4,0)$ as a function of $g_4$ is known for the pure $\phi^4$ theory with $g_6=0$~\cite{Lipatov1977, Zinn-Justin1981}. 
We first review this calculation and then generalize it to $g_6>0$. 
When $g_6=0$, $G_n(g_4, 0) = \sum_k A_k^{(n)} g_4^k$ has a branch cut at $g_4 \in (-\infty, 0)$. 
As shown for the toy model in Eq.~(\ref{coefficient}), the expansion coefficient is related to the branch cut by
\bea
	A_k^{(n)} = \int_{-\infty}^0 \frac{dz}{\pi} \frac{\Im G_n(z,0) }{z^{k+1}}.
\eea

Since the unstable saddle-point solution at $g_4<0$ will contribute to the imaginary part, we use the saddle-point approximation. 
When $g_6 = 0$, the saddle-point equation is
\bea\label{SEq1}
	-\partial^2 \varphi_c+ g_4 \varphi_c^3 = 0.
\eea
Let  $\varphi_c(x) = \frac{\xi(x)}{\sqrt {-g_4}}$, where $\xi$ is the solution of $ \partial^2 \xi+ \xi^3 = 0$ and is independent of $g_4$. In $d=4$ dimensions, there are two solutions: $\xi(x) = \pm \frac{2\sqrt2}{x^2+1}$ (see Appendix~\ref{app:saddlepoint}). Note that the saddle-point equation is scale-invariant in the sense that if $\xi(x)$ is a solution, then $b^{-1}\xi(x/b)$ is also a solution for any rescaling parameter $b$. We will remove this scaling redundancy later on. Because $\int d^d x [(\partial\varphi_c)^2 + g_4 \varphi_c^4] = 0$, the on-shell action is
\bea
	S[g_4,0; \varphi_c] = - \frac{g_4}4 \int d^d x \varphi_c^4 = - \frac{S_0}{4g_4},
\eea
where $S_0 \equiv \int d^dx \xi^4$ is a positive constant independent of $g_4$. In $d=4$ dimensions, $S_0=32\pi^2/3$.

The fluctuations around the saddle-point solution can be expanded in terms of an orthonormal basis $\{\varphi_j(x) \}$, 
\bea
	\phi(x)= \varphi_c(x-x_0) + \sum_{j=1}^\infty c_j \varphi_j(x).
\eea
Here, $x_0$ is an arbitrary point in spacetime. 
Using this basis, one can change the functional integration into a c-number integration. Note that to make the integration variable linearly independent, one requires
\bea
	\int d^d x  \varphi_j(x) \partial_\mu \varphi_c(x) =0, \quad \mu= 1,...,d.
\eea
Moreover, the Jacobian that results from the change of integration variables from the functional measure to $x_0$ and $c_n$ is given by
\bea
	J &=& \left[ \prod_{\mu=1}^d \int d^dx (\partial_\mu \varphi_c)^2 \right]^{1/2} = \left[  -\frac{S_0}{dg_4} \right]^{d/2},
\eea
where we have used the spherical symmetry of the saddle-point solution. The quadratic kernel around the saddle point is
\bea
	M(x,x') &=& \frac{\delta^2 S[g_4,0;\varphi_c]}{\delta \varphi_c(x) \delta \varphi_c(x')}	= \left[ -\partial^2_x - 3 \xi^2(x) \right] \delta(x-x'), \nn \\
\eea
which is independent of $g_4$. 
Note that $\partial_\mu \varphi(x)$ is the eigenfunction of the kernel with zero eigenvalue. 
The above orthonormal basis $\{\varphi_j \}$ can be chosen to be the eigenfunctions of $M(x,x')$.

Putting everything together, the imaginary part of the correlation function under the saddle-point approximation is given by
\bea\label{Gn4}
	 \Im G_n(g_4,0) &=& \frac1{2i} \sum_{\varphi_c} \int J d^d x_0 e^{-S[g_4, 0; \varphi_c]}\int D\varphi_\perp \nn \\
	&& \times \prod_{j=1}^n \varphi(x_j) e^{- \int d^dx d^dx'\varphi_\perp(x) M(x,x') \varphi_\perp(x')} \nn\\
	&=& \left(\frac{S_0}{d} \right)^{d/2} \frac{e^{\frac{S_0}{4g_4}}}{(-g_4)^{d/2+ n/2}}  \frac{F_n(X_n;\xi)}{i(\det M)^{1/2}},
\eea
where $F_n(X_n; \xi)= \int d^d x_0 \prod_j \xi(x_j-x_0)$, $n\in 2\mathbb Z$, and $\varphi_\perp$ denotes the functions orthogonal to $\partial_\mu \varphi$. 
$\sum_{\varphi_c}$ denotes the summation over the two saddle points. 
The expansion coefficient is
\bea\label{phi4A}
	A_k^{(n)} &=& (-1)^k \left(\frac{S_0}{d} \right)^{d/2} \frac{ i F_n(X_n)}{(\det M)^{1/2}}  \nn \\
	&& \times \int_{-\infty}^0 \frac{dz}\pi \frac{e^{\frac{S_0}{4z}}}{(-z)^{k+d/2+n/2+1}} \nn \\
	 &=& (-1)^{k} \frac{1}\pi \Big(\frac{4}d\Big)^{\frac{d}2} \frac{i F_n(X_n;\xi)}{(\det M)^{1/2}}  \Big(\frac{4}{S_0}\Big)^{k+\frac{n}2} \Gamma(k+ \frac{d}2 + \frac{n}2 ). \nn \\
\eea
Because $M(x,x')$ has a unique negative eigenvalue~\cite{Zinn-Justin1981}, $\frac{i}{(\det M)^{1/2}}$ is a positive number.

We now generalize this procedure to the case of $\phi^4 + \phi^6$ theory. We first treat the pure $\phi^6$ theory nonperturbatively and then include perturbative corrections from the $\phi^4$ term. 
The saddle-point equation of a pure $\phi^6$ theory is
\bea\label{SEq2}
	-\partial^2 \phi_c  + g_6 \phi_c^5 = 0.
\eea
The solution is $\phi_c(x) = \frac{\chi(x)}{(-g_6)^{1/4}}$, where $-\partial^2 \chi - \chi^5 = 0$. 
Because $\int d^d x [(\partial\phi_c)^2 + g_6 \phi_c^6] =0$, the on-shell action is
\bea
	S[0,g_6; \phi_c] = - \frac{g_6}3 \int d^d x \phi_c^6 = \frac{S_6}{3\sqrt{-g_6}},
\eea
where $S_6 \equiv \int d^d x \chi^6(x)$ is a positive constant independent of $g_6$.

As before, fluctuations around the saddle-point solution can be expanded in an orthonormal basis $\{\phi_j(x) \}$, 
\bea
	\phi(x)= \phi_c(x-x_0) + \sum_{j=1}^\infty c_j \phi_j(x),
\eea
and $ \int d^d x  \phi_j(x) \partial_\mu \phi_c(x) =0$, for $ \mu= 1,...,d $. In this case, the Jacobian is
\bea
	J &=& \left[ \prod_\mu \int d^dx (\partial_\mu \phi_c)^2 \right]^{1/2} 
	= \left(\frac{S_6}{d\sqrt{-g_6}} \right)^{d/2}.
\eea
The quadratic kernel around the saddle point is
\bea
	\mathcal M(x,x') 
	= \left[ -\partial^2_x - 5 \chi^4(x) \right] \delta(x-x'),
\eea
which is independent of $g_6$. We evaluate the imaginary part of the correlation function using the saddle-point approximation, which yields 
\bea
	\Im G_n(0,g_6) &=& \frac1{2i} \sum_{\phi_c} \int J d^d x_0 e^{-S[\phi_c]}\int D\phi_\perp \nn \\
	&& \times \prod_{j=1}^n \phi(x_j) e^{- \int d^dx d^dx'\phi_\perp(x)\mathcal M(x,x') \phi_\perp(x)} \nn\\
	&=& \left(\frac{S_6}{d} \right)^{d/2} \frac{e^{-\frac{S_6}{3\sqrt{-g_6}}}}{(-g_6)^{d/4+ n/4}}  \frac{F_n(X_n)}{i(\det \mathcal M)^{1/2}}.
\eea
where $F_n(X_n; \chi)= \int d^d x_0 \prod_j \chi(x_j-x_0)$ and $n \in 2 \mathbb Z$. 
The expansion coefficients are
\bea
	&& B_k^{(n)}(0) \nn \\
	&=& (-1)^k \left(\frac{S_6}{d} \right)^{d/2} \frac{ i F_n(X_n)}{(\det \mathcal M)^{1/2}} \int_{-\infty}^0 \frac{dz}\pi \frac{e^{-\frac{S_6}{3\sqrt{-z}}}}{(-z)^{k+d/4+n/4+1}} \nn \\
	 &=& (-1)^{k} \frac{2 }{\pi} \Big(\frac{3}d\Big)^{\frac{d}2}\frac{i F_n(X_n; \chi)}{(\det \mathcal M)^{1/2}}  \Big(\frac{3}{S_6}\Big)^{2k+\frac{n}2} \Gamma\Big(2k+ \frac{d}2 + \frac{n}2 \Big). \nn\\
\eea
The fact that the kernel $\mathcal M$ has a negative eigenvalue can be seen by noting that 
\bea\label{eq:Mchikernel}
	\int d^d x d^d x' \chi(x) \mathcal M(x,x') \chi(x')= -4 \int d^dx \chi^6(x) < 0. \nn\\
\eea
It remains to be shown that there is an odd number of negative eigenvalues. 
We assume this is true, in which case $\frac{i}{(\det \mathcal M)^{1/2}}$ is a real number.

Now we can consider the correction from a finite $\int d^d x g_4 \phi^4$ term. 
The first-order perturbation is given by the correction to the on-shell action,
\bea
	S[g_4, g_6 ; \phi_c] \approx \frac{S_6}{3\sqrt{-g_6}} +  g_4 \int d^d x \phi_c^4 = \frac{S_6}{3\sqrt{-g_6}} - \frac{g_4}{g_6} S_4,  \nn \\
\eea
where $S_4 = \int d^d x \chi^4 $ is a positive constant independent of $g_4$ and $g_6$. 
Then the asymptotic coefficient is
\bea
	&& B_k^{(n)}(g_4) \nn \\
	&=& (-1)^k \left(\frac{S_6}{d} \right)^{\frac{d}2} \frac{ i F_n(X_n;\chi)}{(\det \mathcal M)^{1/2}} \int_{-\infty}^0 \frac{dz}\pi \frac{e^{-\frac{S_6}{3\sqrt{-z}} + \frac{g_4}{z} S_4}}{(-z)^{k+d/4+n/4+1}} \nn \\
	 &=& (-1)^{k} \frac{2 }{\pi} \Big(\frac{3}d\Big)^{\frac{d}2}\frac{i F_n(X_n; \chi)}{(\det \mathcal M)^{1/2}}  \Big(\frac{3}{S_6}\Big)^{2k+\frac{n}2} \Gamma(2k+ \frac{d}2 + \frac{n}2 )  \nn \\
	 && \times \mathcal U\Big(k+ \frac{d}4 + \frac{n}4, \frac12, \frac{S_6^2}{36 S_4} \frac1{g_4} \Big),
\eea
where we have defined
\bea\label{UFunction}
\mathcal U(a,b,c) = c^{a} U(a,b,c),
\eea
and $U$ is Tricomi's (confluent hypergeometric) function. 
One can further expand Tricomi's function to get the expansion coefficients of $g_4^j g_6^k$, as we will do in the next section.

\subsubsection{Beta functions at large orders}
In the previous section, we computed the asymptotic expansion of the $n$-point correlation function of the $\phi^4 + \phi^6$ theory. 
In this section, we use this result to compute the high-order expansion of the beta function.

We define the renormalized couplings as the full $n$-point vertex functions, $\lambda_n = -\Gamma_n(\bar g_{4}, \bar g_{6}; \frac{p^2}{\mu^2})$, where $n=4, 6$, and
$\bar g_4 = g_4$ and $\bar g_6 = g_6 \mu^{2}$ are the dimensionless couplings. Here, $p$ is the momentum at the renormalization point, and $\mu$ denotes the energy scale. From the equations defining the saddle points, Eqs.~(\ref{SEq1}) and~(\ref{SEq2}), we have
\bea
	\xi(x) &=& \int d^d y \Delta_0(x-y) \xi^3(y), \\
	\chi(x) &=& \int d^d y \Delta_0(x-y) \chi^5(y),
\eea
where $\Delta_0(p) = \frac1{p^2}$ is the free propagator. We then have
\bea\label{Fn}
	F_n(X_n; s) &=& \int d^d y F_s(y) \prod_{j=1}^n \Delta_0(x_j-y) ,
\eea
where $s= \xi, \chi$, e.g., $F_\xi(y)= \int d^d x_0  \xi^3(y-x_0)$ and $F_\chi(y)= \int d^d x_0  \chi^5(y-x_0)$. Equation~(\ref{Fn}) converts the correlation function to the vertex function~\cite{Lipatov1977}. 
There is a subtlety for $d=4$ dimensions because the $\phi^4$ potential is scale invariant in this case. If 
$\xi(x)$ is a solution to the saddle-point equation, $ \partial^2 \xi+ \xi^3 = 0$, then it follows that $b^{-1} \xi ( x/b) $ is also a solution for any rescaling factor $b$. We can remove this trivial scaling redundancy by using the following identity:
\bea
	1 = S_0 \int d \log b^2 \delta \Big( \int d^4 x \xi^4(x) \log \frac{x^2}{b^2} \Big),
\eea
in conjunction with the rescaling transformation $ \xi (x) \rightarrow b^{-1} \xi ( x/b) $. 
The $\delta$ function above fixes the scaling of the solution, namely, $\xi(x) = \pm \frac{2\sqrt2}{x^2+1}$. 
These modifications are taken into account by working with the function
\bea\label{F4}
	\tilde{F}_\xi(y)= S_0 \int d\log b^2 \int d^d x_0  b^{-3} \xi^3 \Big( \frac{y-x_0}b \Big).
\eea 
Putting everything together, we get
\bea
	\Gamma_n\Big(\bar g_4, \bar g_6; \frac{p^2}{\mu^2}\Big) 	&=& \sum_k \A_k^{(n)} \bar g_4^k + \B_{j,k}^{(n)} \bar g_4^j \bar g_6^k ,
\eea
and the asymptotic forms of the coefficients are given by
\bea\label{asymptotic}
	\mathcal A_k^{(n)} &=&  \frac{(-1)^{k}}\pi \Big(\frac{4}d\Big)^{\frac{d}2} \frac{i \tilde F_\xi(p)}{(\det M)^{1/2}} \nn \\
	&& \times \Big(\frac{4}{S_0}\Big)^{k+\frac{n}2} \Gamma(k+ \frac{d}2 + \frac{n}2 ), \\
	 \B_{m,k}^{(n)} &=&   \frac{2 (-1)^{k}}{\pi} \Big(\frac{3}d\Big)^{\frac{d}2}\frac{i \mu^2 F_\chi(p)}{(\det \mathcal M)^{1/2}}  \nn \\
	 &&\times \Big(\frac{3}{S_6}\Big)^{2k+\frac{n}2} \Gamma(2k+ \frac{d}2 + \frac{n}2 )  \frac{\partial^m \mathcal U}{m!},
\eea
where $F_s(p), \tilde F_s(p)$ ($s=\xi, \chi$) are the Fourier transforms of $F_s(x), \tilde F_s(x)$.

The function Eq.~(\ref{UFunction}) has the expansion
\bea
	\mathcal U(a,b,c) = \sum_{n=0}^\infty \frac{(a)_n (a+1-b)_n}{n!} (-c)^{-n},
\eea
where $(a)_0\equiv1, (a)_n \equiv a(a+1)...(a+n-1)$. 
The two limits of $\frac{\partial^m \mathcal U}{m!}$ are given by
\bea
	\frac{\partial^m \mathcal U}{m!} \rightarrow \begin{cases} \frac{k^{2m}}{m!} \Big(-\frac{36 S_4}{S_6^2}\Big)^m, \quad &k\gg m \gg 1,\\
	\frac{m! m^{2k}}{(k!)^2} \Big(-\frac{36 S_4}{S_6^2}\Big)^m, \quad & m \gg k \gg 1.
	\end{cases}
\eea
from which we see that the asymptotic behaviors, $\A_k^{(n)}$ and $\B_{k,m}^{(n)}$ for fixed $m$ increase factorially as $k!$, while $\B_{m,k}^{(n)} $ for fixed $m$ increases faster as $(k!)^2$, i.e.,
\begin{widetext}
\bea
\label{A}	\A_k^{(n)} &\propto&  (-1)^k \Big(\frac{4}{S_0}\Big)^{k}  k^{\frac{d}2 + \frac{n}2-1} k!, \quad \quad \quad \quad k \gg 1\\
\label{B}	\B_{m,k}^{(n)} &\propto& \begin{cases} (-1)^{k+m} \Big(\frac{3}{S_6}\Big)^{2k} \Big(\frac{36 S_4}{S_6^2}\Big)^m \frac{k^{\frac{d+n-3}2+2m}}{m!} 4^k (k!)^2, &  k\gg m \gg1 \\
	(-1)^{k+m} \Big(\frac{3}{S_6}\Big)^{2k} \Big(\frac{36 S_4}{S_6^2}\Big)^m k^{\frac{d+n-3}2} 4^k m^{2k} m!, &  m\gg k \gg 1
	\end{cases} 
\eea
\end{widetext}
We are interested in the $\phi^4+ \phi^6 $ theory in four dimensions, the coupling constants $\lambda_4$ and $\lambda_6$ are
\bea
	\lambda_4 &=& \bar g_4 - \sum_{k=2}^\infty \A_k^{(4)} \bar g_4^k -  \sum_{j, k} \B_{j,k}^{(4)} \bar g_4^j \bar g_6^k, \\
	\lambda_6 &=& - \sum_{k=2}^\infty \A_k^{(6)} \bar g_4^k + \bar g_6 -\sum_{j, k} \B_{j,k}^{(6)} \bar g_4^j \bar g_6^k.
\eea
from which we can reexpand $\bar g_n$ as a function of $\lambda_n$. The leading orders are given by
\bea
	\bar g_4 &=& \lambda_4 +  \B^{(4)}_{0,1} \lambda_6 + \O(\lambda_4^2, \lambda_6^2, \lambda_4 \lambda_6), \\
	\bar g_6 &=& \lambda_6 + \O(\lambda_4^2, \lambda_6^2, \lambda_4 \lambda_6).
\eea
The beta functions have the asymptotic expansion,
\bea\label{lambda4} 
	\frac{d\lambda_4}{d\log \mu} &=& - \sum_{k=2}^\infty \partial \A_k^{(4)} \bar g_4^k -  \sum_{j, k} \partial \B_{j,k}^{(4)} \bar g_4^j \bar g_6^k ,  \nn \\
&\approx & - \sum_k \partial \A_k^{(4)}  [\lambda_4^k + (\B^{(4)}_{0,1} \lambda_6)^k] \nn \\
&& - \sum_{j,k } \partial\A_{j+k}^{(4)} \frac{(j+k)!}{j!k!} \lambda_4^j (\B^{(4)}_{0,1} \lambda_6)^k  \nn \\
&& -  \sum_{k} \partial  \B_{j,k}^{(4)} \lambda_4^j \lambda_6^k, \\ 
\label{lambda6} 
	\frac{d\lambda_6}{d\log\mu} &=& - \sum_{k=2}^\infty \partial\A_k^{(6)} \bar g_4^k -  \sum_{j, k} \partial\B_{j,k}^{(6)} \bar g_4^j \bar g_6^k , \nn \\ 
&\approx & - \sum_k \partial\A_k^{(6)}  [\lambda_4^k + (\B^{(4)}_{0,1} \lambda_6)^k] \nn \\
&& - \sum_{j,k } \partial\A_k^{(6)} \frac{(j+k)!}{j!k!} \lambda_4^j (\B^{(4)}_{0,1} \lambda_6)^k  \nn \\
&& -  \sum_{k} \partial\B_{j,k}^{(6)} \lambda_4^j \lambda_6^k.
\eea
where $\partial \A = \frac{\partial}{\partial \log \mu}  \A$.

Equations~(\ref{A}),~(\ref{B}),~(\ref{lambda4}) and~(\ref{lambda6}) are the main results of this section. 
We see from these results that the inclusion of the $\phi^6$ term, which is allowed by symmetry in condensed matter systems (and, in any case, is automatically generated in the theory from the $\phi^4$ term),  dramatically changes the asymptotic behavior of the beta function for $\lambda_4$. 
On the one hand, if $\lambda_6 \sim \lambda_4$, the asymptotic coefficient of the beta function for $\lambda_4$ changes qualitatively from $k!$ to $(k!)^2$: 
\bea
	 \B_{0,k}^{(4)} &\propto&  \Big(\frac{6}{S_6}\Big)^{2k}  k^{\frac{d+1}2}  (k!)^2,  \quad k\gg 1.
\eea
Thus, we see that the inclusion of $\lambda_6$ completely changes the asymptotic behavior of the beta function for $\lambda_4$. In light of Eq.~\eqref{eq:Landaupoletest}, this will in turn change the fate of the Landau pole. 
On the other hand, even if $\lambda_6 \ll \lambda_4$, as long as the $\phi^6$ term is nonzero $\lambda_6 >0$, the asymptotic form of the second term in Eq.~(\ref{lambda4}) is comparable to the first term of the pure $\lambda_4$, i.e.,
\bea
	\partial\A_{1+k}^{(4)} &\propto& \Big(\frac{4}{S_0}\Big)^{k}  k^{\frac{d}2 + 1} k!, \quad k\gg1,
\eea
which can also qualitatively change the behavior of the beta function compared to the pure $\phi^4$ theory. 
In particular, including the $\phi_6$ term will result in a new asymptotic series, so even if the pure $\lambda_4$ series does not end up with a Landau pole, it is possible that including the $\phi^6$ term will lead to a Landau pole.

We are not aware of a technique to resum an asymptotic series with multiple variables, but it is still useful to illustrate how the Borel sum of $\lambda_4$ is affected by the $\phi^6$ potential. Notice that a Borel transform~\cite{Borel1928} will take the function $B(g)$ with a branch cut at $-\alpha$, 
\bea
	B(g) = \Big(1+  \frac{g}\alpha \Big)^{-\beta-1} = \sum_k B_k g^k, \\
	B_k = (-\alpha)^{k} \frac{(k+\beta)!}{k!\beta!} \rightarrow (-\alpha)^k k^{\beta},
\eea
to a series $W(g) = \sum_k W_k g^k $ whose asymptotic coefficient is given by
\bea\label{Wk}
	W_k \rightarrow (-\alpha)^k k^{\beta} k!.
\eea

Comparing Eq.~(\ref{Wk}) to Eqs.~(\ref{A}) and~(\ref{B}), a naive Borel sum without considering the $\phi^6$ potential is given by 
\bea\label{Borel0}
	B^{(0)}(\lambda_4) \sim \Big(1+  \frac{S_0}4 \lambda_4 \Big)^{-4}.
\eea
The presence of $\lambda_6$ leads to a different (naive) Borel sum that at $k$-th order has the form 
\bea\label{BorelK}
	B^{(k)}(\lambda_4) \sim \lambda_6^k \Big(1+  \frac{S_6^2}{36S_4} \lambda_4 \Big)^{-2k-1}.
\eea
It is apparent that these Borel sums lead to different asymptotic behaviors. 
Nevertheless, we need to emphasize that instead of having multiple Borel sums, like Eqs.~(\ref{Borel0}) and~(\ref{BorelK}), for each power in $\lambda_6$, it is more appropriate to treat the asymptotic expansions of $\lambda_4$ and $\lambda_6$ on equal footing, because $\lambda_6$ has a vanishing radius of convergence. 

\section{Dirac fermions with Coulomb interactions\label{DFCI}}

In this section, we consider $d$-dimensional Dirac fermions with Coulomb interactions. 
Ultimately, we are interested in Dirac fermions in graphene, where $d=2+1$, with Coulomb interactions. 
This is the condensed matter analog of QED in the context of Landau poles. 
To have a local quantum field theory, one can implement a Hubbard-Stratonovich transformation to decouple the Coulomb interaction by introducing a scalar field. 
The theory under consideration is then given by \cite{Son2007}
\bea\label{coulombScalar}
	S &=& \int d^d x ( \bar \psi \gamma^0 \partial_0 \psi + v \bar \psi \sum_{i=1}^{d-1} \gamma^i \partial_i \psi + i g \phi \bar \psi \gamma^0 \psi) \nn \\
	&&+ \frac1{2} \int d^4 x \sum_{i=1}^3(\partial_i \phi)^2,
\eea
where $\psi$ ($\phi$) refers to the four-component Dirac fermion (the Hubbard-Stratonovich scalar field), and $v$ ($e$) refers to the Fermi velocity (the coupling strength). $\gamma^\mu$, $\mu=0,...,d-1$, denotes the $4\times4$ $\gamma$ matrices. 
The effective interaction strength is given by $\alpha = \frac{N g^2}{16v}$ (also known as the fine structure constant), where $N=2$ is the spin degeneracy, and $g^2=\tfrac{2e^2}{(1+\varepsilon)\epsilon_0}$, where $\varepsilon$ is the background dielectric constant (which in general can be $>1$ in solid state systems), and $\epsilon_0$ is the vacuum permittivity. (Note that the definition of $\alpha$ used here contains an additional factor of $\pi N/4$ compared to that used in Ref.~\cite{SDS2014}.)

As in the previous analysis of critical quantum magnets, we calculate the behavior of the beta function of the fine structure constant at both small and large orders of perturbation theory. Small-order terms can be calculated using either the standard perturbation theory in $\alpha$ or a large-$N$ analysis in which $1/N$ is the expansion parameter. 
Here, we review the results obtained previously for both methods~\cite{Son2007,Hofmann2014,SDS2014}, and we discuss their implications for the Landau pole problem. To our knowledge, the large-order terms of the asymptotic series in $\alpha$ had not been computed prior to the present work. Here, we obtain these by first integrating out the Dirac fermion and then using a saddle-point approximation similar to the one employed in the previous section on quantum magnets.

\subsection{Small orders}
\begin{figure}
\subfigure[]{\label{QED_Fig1}
	\includegraphics[width=2cm]{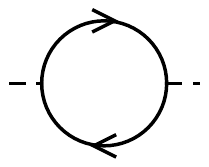}} \\
\subfigure[]{\label{QED_Fig2}
	\includegraphics[width=2cm]{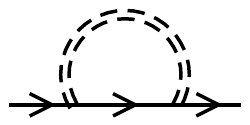}} \quad
\subfigure[]{\label{QED_Fig3}
	\includegraphics[width=1.8cm]{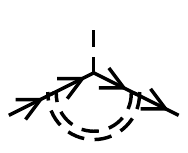}} \quad
\subfigure[]{\label{QED_Fig4}
	\includegraphics[width=2cm]{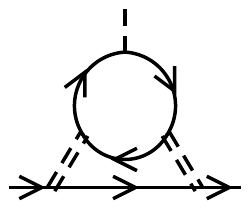}}
	\caption{(a) The polarization of the scalar field. The solid line (single-dashed) represents the Dirac fermion propagator (the bare scalar propagator). (b-d) The 1/N order Feynman diagrams. The double-dashed line represents the screened scalar propagator.}
\end{figure}

The two-loop beta equation was calculated in Ref.~\cite{SDS2014} for the case of graphene (where $d=2+1$ and we have $N=2$ flavors of four-component Dirac fermions), 
\bea\label{2loopRG}
	\frac{d\alpha}{d \log\mu} = f_1 \alpha^2 + f_2 \alpha^3,
\eea
where $\alpha = \frac{g^2}{8v}$, and $f_1= \frac1{2\pi}$, and $f_2= \frac{2\log2-8/3}{\pi^2}$. 
Importantly, we see that $f_2<0$, so when $\alpha$ becomes large, the negative $\alpha^3$ term dominates and prohibits $\alpha$ from becoming larger.
Therefore, the Landau pole is replaced by a fixed point at $\alpha^\ast = - \frac{f_1}{f_2}$. 
This is similar to the $\phi^4 + \phi^6$ theory in the previous section. 
(A similar strong-coupling fixed point is found in the direct nonperturbative numerical simulation of the usual (3+1)-dimensional QED on the lattice in Ref.~\cite{Stuben1998}, where the Landau pole was found to lie in a region of the parameter space made inaccessible by a strong-coupling chiral symmetry-breaking transition.) 
However, evidence for the two-loop critical point in Eq.~(\ref{2loopRG}) has not been observed in graphene experiments, suggesting that the second-order perturbation theory in $\alpha$ may not be reliable, as was argued in Ref.~\cite{SDS2014}. 
In particular, the existence of such a possible strong-coupling fixed point preempting the Landau pole may necessitate a stability analysis going to higher orders. 
For example, if the two terms in Eq.~\eqref{2loopRG} were the leading terms of a geometric series, then the result to all orders would be~\cite{SDS2014}
\bea\label{2loopB}
	\frac{d\alpha}{d \log\mu} = f_1^2 \frac{\alpha^2}{f_1- f_2 \alpha},
\eea
which gives rise to $\beta(\alpha \ll 1) \propto \alpha^2$ at small $\alpha$, and $\beta(\alpha \gg 1) \propto \alpha $ at large $\alpha$. This time there is no ultraviolet fixed point in the theory defined by Eq.~(\ref{2loopB}), and the coupling grows indefinitely with energy scale. There is still no Landau pole, but the mechanism by which it is avoided is quite different. 
A one-loop theory with just the first term in Eq.~(\ref{2loopRG}) clearly has a Landau pole whereas neither the full Eq.~(\ref{2loopRG}) with both terms or the resummed theory of Eq.~(\ref{2loopB}) has a Landau pole (albeit for very different reasons). 
This illustrates how leading-order one-loop perturbative calculations can be misleading when it comes to determining the existence of a Landau pole in a given theory. Similar trends can occur in large-$N$ calculations, as we now discuss in the case of graphene.

Besides expanding in the coupling strength, it is often useful to consider a large-$N$ expansion (in the number of fermion flavors rather than a perturbative expansion in the coupling constant) instead, especially when the bare coupling parameter is not small. This is the situation for free-standing graphene, where $\alpha_\text{bare}\approx 3.4$, suggesting that an expansion in powers of $1/N$ with $N=2$ would be more reliable than an expansion in powers of $\alpha_\text{bare}$ (simply by virtue of the fact that $\alpha>1$ and $1/N<1$). 
This was confirmed in Ref.~\cite{Hofmann2014}.
In the following, we review the large-$N$ calculation for graphene and show that a Ward identity guarantees that only one parameter, the Fermi velocity, renormalizes.
Moreover, we show that the beta function obtained from the leading-order large-$N$ calculation does not exhibit a Landau pole. We believe that the same remains true for the next-to-leading-order theory in $1/N$ also, which is likely to produce only small quantitative corrections to the leading order $1/N$ theory.

The leading large-$N$ diagram shown in Fig.~\ref{QED_Fig1} gives rise to the dressed boson propagator,
\bea
	D(q) = \Big( 2 |\vec q| + \frac{g^2 N}8 \frac{|\vec q|^2}{\sqrt {q^2}} \Big)^{-1},
\eea
where $q^2 = q_0^2 + |\vec q|^2$.
The Feynman diagrams at the next order are shown in Figs.~\ref{QED_Fig2}-\ref{QED_Fig4}. Before calculating these diagrams, it is worth noting that there is a residual gauge invariance in Eq.~(\ref{coulombScalar}) corresponding to the transformation 
\bea
\psi \rightarrow e^{i \theta(\tau)} \psi, \quad \phi \rightarrow \phi - \frac{1}g \partial_\tau \theta(\tau),
\eea
where $\theta$ is an arbitrary function of $\tau$. 
This residual gauge symmetry leads to the Ward identity,
\bea
	p_0 \Gamma_3(p_0, \vec p; q_0, \vec q) = i g [ G(q_0 + p_0, \vec q)^{-1} - G(q_0 , \vec q)^{-1} ],
\eea
where $\Gamma_3$ is the vertex function, and $G^{-1}$ is the inverse fermion propagator. 
This Ward identity is similar to the Ward identity in standard QED where the photon is dynamical, except that here it only involves the temporal component. 
The Ward identity relates the charge renormalization factor to the anomalous dimension of the scalar field. 

While the Dirac fermion is confined to a two-dimensional plane, the scalar field in Eq.~(\ref{coulombScalar}) can propagate in the full three-dimensional space. 
Thus, the backreaction from the Dirac fermion is restricted within the two-dimensional plane, and so it does not lead to an anomalous dimension for the scalar field. 
Technically, this is because the scalar propagator is nonlocal in the plane $D^{-1}_0(q) \sim |\vec q|$, and thus it does not receive corrections. 
As a result, according to the Ward identity, the charge does not renormalize. 
The only parameter in Eq.~(\ref{coulombScalar}) that does renormalize is the Fermi velocity $v$, or equivalently, the dynamical exponent $z = 1 +\gamma_v$, where $\gamma_v$ is the anomalous dimension of Fermi velocity. 
Note that this is not true for the theory in 3+1 dimensions, because the scalar field can have a finite anomalous dimension in that case, leading to two independent parameters in 3+1 dimensions. 

In light of the above discussion, the only Feynman diagram we need to evaluate is Fig.~\ref{QED_Fig2}, because the vertex correction can be inferred from the self-energy correction by the Ward identity. 
Using the results from Ref.~\cite{Son2007}, we get the fermion self-energy,
\bea
	\Sigma(p) = \frac4{N\pi^2} [f_0(\alpha) p_0 \gamma^0 + f_1(\alpha) \vec p \cdot \vec \gamma ] \log{\mu}.
\eea
where 
\bea
 f_1(\alpha) = \begin{cases} - \frac{\sqrt{1-\alpha^2}}{\alpha} \arccos \alpha -1 + \frac\pi{2\alpha}, & \alpha<1 \\
		\frac{\sqrt{\alpha^2-1}}{\alpha} \log(\alpha + \sqrt{\alpha^2-1}) -1 + \frac{\pi}{2\alpha}, & \alpha>1
		\end{cases} \nn\\
		\\
f_0(\alpha)
	= \begin{cases} - \frac{2-\alpha^2}{\alpha\sqrt{1-\alpha^2}} \arccos \alpha -2 + \frac{\pi}\alpha, & \alpha<1 \\
		\frac{\alpha^2-2}{\alpha\sqrt{\alpha^2-1}} \log(\alpha + \sqrt{\alpha^2-1}) -2 + \frac{\pi}\alpha,  & \alpha>1
		\end{cases}	\nn \\
\eea
from which we can extract the dynamical exponent, $1-z= \frac4{\pi^2 N} [f_1(\alpha) - f_0(\alpha)]$, and the beta equation,
\bea
	\frac{\partial \alpha}{\partial \log \mu} &=&  (1-z) \alpha.
\eea
We can expand the beta function at small and large $\alpha$:
\bea\label{largeNRG}
	\beta(\alpha) = \begin{cases} \frac1N\Big( \frac1{\pi} \alpha^2 - \frac{8}{3\pi^2} \alpha^3+ O(\alpha^4) \Big) , \quad & \alpha \ll 1\\
	\frac1N\Big( \frac4{\pi^2} \alpha- \frac{4}{\pi} + O(\alpha^{-1}) \Big) . \quad  & \alpha \gg 1
	\end{cases}
\eea
It is apparent that there is no Landau pole because $\beta(\alpha \gg 1) \propto \alpha$. Instead, the coupling diverges as the energy scale goes to infinity. However, if we expanded perturbatively in $\alpha$ (and also in $1/N$), 
and retained terms up to order $\alpha^3$, we would erroneously conclude that there is a fixed point at $\alpha=3\pi/8$. 
This is similar to the fixed point [c.f. Eq.~(\ref{2loopRG})] we found from our original perturbation theory in $\alpha$ above. Of course, if we instead stopped at order $\alpha^2$, we would obtain $\beta(\alpha \ll 1) = \frac{\alpha^2}{\pi N} $, and we would conclude that there is a Landau pole as in the leading-order perturbative 1-loop theory [i.e., Eq.~(\ref{2loopRG}) with just the first term on the right-hand side]. Thus, we see that truncating the series at a low order in $\alpha$ can produce misleading results about the Landau pole issue. 
This is again an indication that the Landau pole in graphene, which is apparent in the 1-loop perturbative RG theory as in the original QED context \cite{Landau1954}, may be an artifact of perturbation theory.

\subsection{Large orders}

Now we turn to the large-order expansion of the beta function. 
Because of the fermionic nature of Dirac fields, the saddle-point analysis is not directly applicable. 
We can circumvent this problem by integrating over the Grassmanian field, which results in a functional determinant. 
In the strong-coupling limit, the determinant can be evaluated using the quasilocal approximation (see Appendix~\ref{app:det}), which then yields an effective action involving just the scalar field~\cite{Zinn-Justin1981, Parisi1977a}. 
In the following, we start from this effective action and apply a saddle-point analysis to obtain the large-order expansion of the beta function.

The effective action is given by~\cite{Zinn-Justin1981, Parisi1977b, Parisi1978}
\bea
\label{detEff}	
S &=& \int d^4 x \frac12 (\nabla \phi)^2 + \int d^d x \alpha_d |\phi|^d,
\eea
where $\alpha_d=  \frac{2\Gamma(-d/2)}{(4\pi)^{d/2}} \frac{N}{v^{d-1}} (g^2)^{d/2}$. Notice that it involves a different combination of $g$ and $v$ because of the different approximation we implement in the large-order series. However, in $2+1$ dimensions, since from the previous subsection we know that the RG correction to the charge $g$ vanishes and all RG effects come from the renormalization of the Fermi velocity $v$, $\alpha_3 =  \frac{1}{3\pi} \frac{N}{v^{2}} |g|^3$ is directly connected to $\alpha$ in the weak-coupling expansion approach. 
The large-order perturbation series for Dirac fermions with Coulomb interactions is equivalent to that of the effective scalar field theory given by Eq.~(\ref{detEff}). 
The potential term in Eq.~(\ref{detEff}) leads to a branch cut in the $n$-point correlation function for $\alpha_d \in (-\infty,0)$. This allows us to use similar techniques as before to evaluate the asymptotic expansion of the correlation function.

The equation of motion that follows from Eq.~(\ref{detEff}) is
\bea
	- \nabla^2 \phi_c + d \alpha_d \text{sign}(\phi_c) |\phi_c|^{d-1} \delta^{(4-d)}(z_\perp)= 0,
\eea
where $z_\perp$ denotes the coordinates perpendicular to the spacetime where the Dirac fermions live. 
For instance, $z_\perp = z$ is the direction perpendicular to the plane where the Dirac fermions are confined in the case of $d=2+1$ dimensions (graphene). 
The equation of motion is solved by $\phi_c = |d \alpha_d|^{- \frac1{d-2}}\eta$, where $\eta$ is a solution to the differential equation $-\nabla^2 \eta - \text{sign}(\eta) |\eta|^{d-1} \delta^{(4-d)}(z_\perp) =0 $. 
Note that here $\alpha_d < 0$ since we want to apply the saddle-point technique.
The on-shell action is
\bea
	S_\text{on-shell} =  
	\frac{d-2}{2d} |\alpha_d|^{- \frac2{d-2}} S_d,
\eea
where $S_d =  \int d^d x d^{- \frac{2}{d-2}}  |\eta|^d $ is a positive constant independent of $\alpha_d$. 

At each time $\tau$, the field configuration can be expressed by an orthogonal basis $\varphi_j (\vec x)$
\bea
	\phi(\tau, \vec x) = \phi_c( \vec x - \vec x_0) + \sum_j f_j(\tau) \varphi_j( \vec x).
\eea
We can change the functional integration measure to $\vec x_0$ and $f_j$. 
The Jacobian associated with the change of variables is given by
\bea
	J &=& \left[ \prod_i \int d^4 x (\partial_i \phi_c)^2 \right]^{1/2} 
	= \left[ \frac{S_d}{3|\alpha_d|^{\frac2{d-2}} } \right]^{3/2},
\eea
and the quadratic kernel around the saddle-point solutions is
\bea
	&&\mathcal M(x,x') = \frac{\delta^2 S[\phi_c]}{\delta \phi_c(x) \delta \phi_c(x')} \nn \\
	&&= \left[ -\nabla^2 - (d-1) |\eta|^{d-2} \delta^{(4-d)}(z_\perp) \right] \delta^{(4)}(x-x').
\eea
It is straightforward to show that this kernel has a negative eigenvalue by projecting it onto $\eta$ in a manner analogous to Eq.~\eqref{eq:Mchikernel}. We again assume that there is an odd number of negative eigenvalues. 

We are interested in the $n$-point correlation function,
\bea
	G_n(\alpha_d;X_n) = \int D\phi \prod_{j=1}^{n}\phi(x_j) e^{-S}.
\eea
In the saddle-point approximation, the imaginary part of this correlation function is
\bea
	 \Im G_n(\alpha_d)
	&=& \frac1{2i} \sum_{\phi_c} \int J d^{3} x_0 e^{-S[\phi_c]}\int D\phi_\perp \nn \\
	&& \times \prod_{j=1}^n \phi(x_j) e^{- \int d^dx d^dx'\phi_\perp(x)\mathcal M(x,x') \phi_\perp(x)} \nn\\
	&=& \frac1{d^{\frac{n}{d-2}}} \Big( \frac{S_d}{3} \Big)^{\frac{3}2} \frac{F_n(X_n; \eta) }{i (\det \mathcal M)^{1/2}} \frac{\exp \Big[ \frac{(2-d)S_d}{2d(-\alpha_d)^{\frac{2}{d-2}}}\Big] }{(-\alpha_d)^{\frac{n+3}{d-2}}},  \nn \\
\eea
where $F_n(X_n; \eta) = \int d^3 x_0 \prod_{j=1}^n \eta(\vec x_j - \vec x_0)$. 
Because the effective action is scale-invariant, i.e., invariant under $\eta(x) \rightarrow b^{-1} \eta(x/b)$, the scaling freedom needs to be fixed in the path integral. 
Using a similar method to what is shown in Eq.~(\ref{F4}), we arrive at $\tilde F_n(X_n; \eta) = S_d \int d \log b^2 \int d^3 x_0 \prod_{j=1}^n b^{-1} \eta( \frac{\vec x_j - \vec x_0}b)$. 
The expansion coefficient is then
\bea\label{DiracCoefficient}
	  A^{(n)}_k(z) 
	&=& (-1)^k \frac1{d^{\frac{n}{d-2}}} \Big( \frac{S_d}{3} \Big)^{\frac{3}2} \frac{i \tilde F(X_n; \eta)}{(\det \mathcal M)^{1/2}} \nonumber\\
	&& \times \int_{-\infty}^0 \frac{dz}{\pi}\frac{\exp \Big[ \frac{(2-d)S_d}{2d}(-z)^{- \frac{2}{d-2}}\Big] }{(-z)^{k+1+\frac{n+3}{d-2}}} \nonumber\\
	&=& (-1)^k \frac{d-2}{2\pi d^{\frac{n}{d-2}}} \Big( \frac{S_d}{3} \Big)^{\frac{3}2} \frac{i F(X_n; \eta)}{(\det \mathcal M)^{1/2}} \nonumber\\
	&& \times \Big( \frac{2d}{(d-2)S_d}\Big)^{\frac{d-2}2 k + \frac{3+n}2} \Gamma\Big( \frac{d-2}2 k + \frac{n+3}2 \Big). \nn\\
\eea
Comparing Eq.~(\ref{DiracCoefficient}) to Eq.~(\ref{phi4A}), we obtain the large-order expansion of the beta equation:
\bea
	\frac{d\alpha}{d \log \mu} &=& \sum_k \C_k \alpha_d^k, \\
	\C_k &\propto& (-1)^k\Big( \frac{2d}{d-2} \frac1{S_d}\Big)^{\frac{d-2}d k} k^{\frac{n+1}2} \Big(\frac{d-2}2 k\Big)!.
\eea
The $k$-th order expansion coefficient of the beta function is proportional to $(\frac{d-2}2 k)!$. 
It is less than the asymptotic expansion coefficient $\sim k!$ in the pure $\phi^4$ theory in $d<4$ dimensions. 
In the $(2+1)$-dimensional Dirac fermion theory with Coulomb interactions, the asymptotic coefficient is proportional to $(\frac{k}2) !$. Since every term in the asymptotic expansion is bounded by that of a pure $\phi^4$ model, we can conclude that if there is no Landau pole in pure $\phi^4$ theory, then there is no Landau pole in the Dirac fermion theory with Coulomb interactions. This is a direct consequence of the fact that the existence of a Landau pole is fully determined by the asymptotic behavior of the beta function, as shown in Eq.~\eqref{eq:Landaupoletest}.

\section{Discussion and conclusion \label{DD}}
Since the seminal works of Wilson~\cite{Wilson1975, Wilson1971}, it has been understood that quantum field theories are essentially effective descriptions of the low-energy and long-wavelength behavior of an underlying physical system, and the quest for an axiomatic foundation of quantum field theories, fashionable during the 1960s, is futile and unnecessary.  Such an effective field theory description is of course particularly germane in condensed matter systems where the existence of a physical lattice imposes a real high-energy short-distance cutoff on any continuum description. 
It is remarkable that such theories can give quantitatively accurate predictions over a wide variety of systems and phenomena, ranging from Kosterlitz–Thouless transitions in Josephson-junction arrays to the tricritical point where water and gas can no longer be distinguished. In the context of particle physics, the Landau pole issue is often viewed as purely academic, because when a pole appears, it is typically at an incredibly high energy scale, and one normally assumes that even if the pole is a real feature of the field theory, the theory itself will likely become invalid by the time this energy is reached. It is difficult to make these statements precise because, in the particle physics context, effective field theories capture the low-energy limit of a system whose high-energy, microscopic degrees of freedom are often unknown. 
In addition, the continuum is presumably real in particle physics since there is no underlying physical lattice providing a short-distance cutoff.  Appealing to the possibility of a theory with the Landau pole becoming inapplicable at high energies (where the pole presumably lies) because some other theories may control the physics at the higher energy scale is not aesthetically pleasing because the disturbing question still remains about the existence of the original theory with the Landau pole (e.g., QED):  Is it a well-defined interacting theory or is it trivial?

The Landau pole problem becomes a bit more concrete in condensed matter systems, because the limitations of such theories are typically known and determined by lattice-scale cutoffs. This raises interesting questions: Is it possible for a Landau pole to occur below the lattice-scale cutoff? What would be the experimental implications of this? The effective field theory point of view makes the Landau pole existence question rather subtle though. The fact that all symmetry-allowed terms can become important in the high-energy regime of an effective field theory makes it particularly challenging to determine if a Landau pole arises. Because the Landau pole is by definition a high-energy phenomenon, it does not matter whether these terms influence the low-energy properties of the theory or not. In the critical quantum magnet described by the $\phi^4$ theory, $\phi^{2n}$ ($n>2$) potentials are present without question. 
Although irrelevant to the long-wavelength critical phenomena, the appearance of a $\phi^6$ potential on top of the $\phi^4$ theory can significantly change the fate of the Landau pole at high energies (which is a short-distance rather than a long-wavelength phenomenon), as we have shown. 
On the one hand, at small orders of the beta function, we find that the Landau pole is removed and replaced by a new ultraviolet fixed point because of the coupled RG flow of $\lambda_4$ and $\lambda_6$. 
On the other hand, at large orders, the asymptotic expansion of the beta function gets a large contribution from $\lambda_6$, which can also alter the fate of the Landau pole. What happens if other $\phi^{2n}$ terms are also included in the analysis? Ultimately, it may have to be decided by experiments that probe strong-coupling, high-energy regimes to finally settle the Landau pole issue in condensed matter systems. 

We note that there is nothing in principle ruling out the possible existence of a Landau pole in condensed matter systems at energy scales well below the ultraviolet lattice cutoff scale, making the issue relevant both theoretically and experimentally. In fact, rough estimates suggest that this may indeed be the case for systems currently under investigation. The well-studied three-dimensional antiferromagnet TlCuCl$_3$ features an O(3) quantum phase transition realized by tuning the pressure. From Ref.~\cite{Scammell2015}, the quartic interaction strength at the quantum critical point of TlCuCl$_3$ is estimated to be around $\lambda_4 \approx 0.23/4\pi$ at $1$ meV, and accordingly the one-loop calculation predicts that the Landau pole occurs at $3.5$ meV, which is far below the lattice scale. For comparison, the measured magnon dispersion at zero pressure (in the disordered phase) reaches as high as $7$ meV~\cite{Sigrist2002}, well above the predicted Landau pole energy. In the ordered phase, the reported gap of the longitudinal excitation reaches about $1.2$ meV~\cite{Boehm2008}. These facts imply that the predicted Landau pole is within the reach of experiments. A possible signature of a Landau pole could come from the fact that a diverging quartic interaction should cause a strong decay of the longitudinal mode to transversal modes in the ordered phase. This in turn suggests that the decay width should exhibit a significant enhancement if there is a Landau pole. However, existing data shows no evidence of such an increase in the line-width~[cf. Ref.~\cite{Sushkov2011}]. On the other hand, there has by no means been an exhaustive search, and a systematic experimental survey at energies approaching the lattice scale is required to settle this issue. We believe that experiments in quantum magnets looking for signatures of Landau poles are needed given that the existence of an infinite number of symmetry-allowed field operators (i.e. all the $\phi^{2n}$ terms) in the theory make a decisive theoretical conclusion impossible.

We also examined a second type of condensed matter system (namely, graphene) searching for Landau poles: Dirac fermions in 2+1 dimensions with Coulomb interactions. This is the QED analog of the Landau pole (albeit in two spatial dimensions). We argued with the help of a Ward identity that the renormalization group flow is controlled by a single parameter~\cite{SDS2014}, the fine structure constant $\alpha \sim \frac{g^2}v$ (or equivalently the Fermi velocity). 
Both perturbation theory in $\alpha$ and the large-$N$ expansion to leading order suggest that the Landau pole is absent in graphene. 
However, keeping only the first few orders in perturbation theory is insufficient to address the issue. 
Therefore, we also evaluated the coefficients of high-order terms in the asymptotic series using a nonperturbative approach adapted from the Lipatov's method. 
Similarly to the case of relativistic fermions with Yukawa type interactions~\cite{Parisi1977a, Zinn-Justin1981}, the asymptotic coefficient is, for $d<4$ dimensions, bounded by that of the four-dimensional pure $\phi^4$ theory, and consequently graphene is free of Landau poles if the pure $\phi^4$ theory does not manifest a Landau pole. 
Moreover, the knowledge of the asymptotic series combined with a few small-order coefficients can be used as the input into a resummation technique that could potentially lead to a resolution of the Landau pole problem that is independent of pure $\phi^4$ theory in this particular context ~\cite{Suslov2001, Suslov2008}. We leave this technically demanding calculation to future work.

Finally, we point to another direction for possible future investigations. As is apparent from the form of the fine structure constant, small velocities can yield large coupling strengths. It is worth noting that the Fermi velocity at the charge neutrality point in twisted bilayer graphene near the magic angle is extremely small~\cite{MacDonald2011, Cao2018a, Cao2018b, SDS2019}. This suggests that this system could be a particularly interesting place to explore the Landau pole problem, assuming the continuum Dirac description is still valid here. 
We can estimate the Landau pole energy scale using the result for the one-loop beta function, which yields $E_L \approx \hbar v_F \sqrt{n} e^{2\pi/\alpha}$, where $n$ is the carrier density and $\alpha$ is the effective coupling (fine structure constant). Typical values for these quantities are $v_F \approx 10^8$ cm/s and $n\approx 10^{12}$ cm$^{-2}$. For graphene grown on a BN substrate where $\alpha \approx 0.4 \pi/2$, $E_L \approx 10^3$ eV, which is several orders of magnitude larger than the lattice cutoff. On the other hand, for suspended graphene where $\alpha \approx 2.2 \pi/2$, the Landau pole energy is around $E_L \approx 0.4$ eV, which is comparable to the lattice scale. Interestingly, for twisted bilayer graphene on hBN where $\alpha \approx 10 \pi/2$, the Landau pole energy is further suppressed to $E_L\approx 98$ meV. This suggests that the Landau pole could be very likely within the reach of experiments. A possible signature could come from the fact that the Fermi velocity should be strongly suppressed as the Landau pole energy scale is approached due to the fact that the velocity gets renormalized and scales as $1/\alpha$. This in turn could lead to interaction-enhanced dispersion flattening and strong effective couplings that could persist as the system is tuned slightly away from a magic angle. In fact, it was reported in Ref.~\cite{SDS2019} that the running coupling constant extracted from experiments is not consistent with a one-loop calculation, indicating that intriguing strong-coupling effects may be taking place in twisted bilayer graphene. Such strong interaction effects, in light of our discussion, may provide important insights on the Landau pole problem as it pertains to Dirac fermions. On the other hand, if a small gap is opening up at the Dirac point in the twisted bilayer graphene, then the Dirac description fails at low energies, and the Landau pole issue becomes moot.  

It may be useful to emphasize that the detailed analysis of Ref.~\cite{SDS2019} focusing on the twisted bilayer graphene experiments of Refs.~\cite{Cao2018a, Cao2018b} and Ref.~\cite{Young2019} came to the conclusion that the measured effective mass and Fermi velocity near the Dirac cone of low-angle twisted bilayer graphene agrees with strong-coupling nonperturbative theories such as the resummed Borel-Pad\'e perturbation series and the $1/N$ expansion, while disagreeing very strongly with the one-loop perturbative RG theory.  In particular, the one-loop theory predicts a very large renormalization of the effective mass and the Fermi velocity for the large ($\alpha \sim 10\pi/2$) effective fine structure constant in the system, which is simply not observed experimentally.  If these preliminary experimental results hold in future measurements in flat band twisted bilayer graphene, where the effective interaction strength is very large, one inevitable conclusion is that the Landau pole as inferred from the running of the coupling implied by the one-loop perturbative RG theory does not exist (and is purely an artifact of the one-loop theory) since the one-loop theory seems unable to quantitatively describe the running of the coupling at large coupling. These preliminary measurements should be repeated in future experiments for a definitive resolution of the question of Landau pole in graphene since its implications extend far beyond graphene all the way to QED, where $\alpha \sim 1/137$, and the RG running of the coupling all the way to the Landau pole in laboratory experiments is manifestly impossible. 
Careful experimental investigation of twisted bilayer graphene near the Dirac point may finally shed light on the 80-year-old question of the existence or not of Landau poles in quantum electrodynamics.
\\

\section*{Acknowledgements}
S.-K.J. is supported by the Simons Foundation through the It from Qubit Collaboration. S.D.S. is supported by the Laboratory for Physical Sciences and the Microsoft  Corporation. E.B. acknowledges support from the National Science Foundation (DMR-1847078).

\appendix

\section{Perturbative renormalization group analysis of $\phi^4 + \phi^6$ theory} \label{app:RG}
\begin{figure}[b]
\subfigure[]{
	\includegraphics[width=1.8cm]{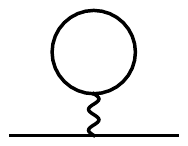}} \quad
\subfigure[]{
	\includegraphics[width=2cm]{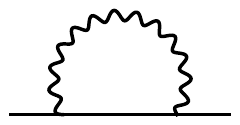}}
\caption{\label{2P}Feynman diagrams contribute to the beta equation of the polarization. The solid line represents the boson propagator of the same index, and the wavy line represent short-range interactions ($\phi^4$ and $\phi^6$ vertices).}
\end{figure}

\begin{figure}
\subfigure[]{
	\includegraphics[width=1.8cm]{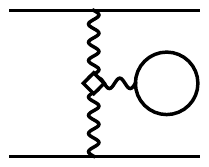}} \quad
\subfigure[]{
	\includegraphics[width=2cm]{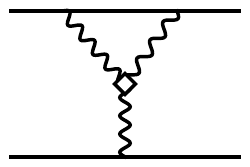}} \quad
\subfigure[]{
	\includegraphics[width=1.8cm]{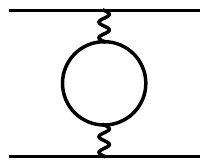}} \quad
\subfigure[]{
	\includegraphics[width=2cm]{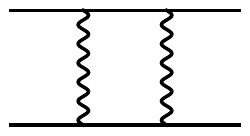}} \quad
\subfigure[]{
	\includegraphics[width=2cm]{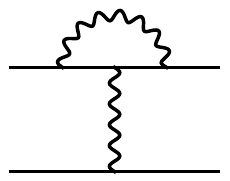}} 
\caption{\label{4P}Feynman diagrams contribute to the beta equation of the four-point vertex. }
\end{figure}

\begin{figure}
\subfigure[]{
	\includegraphics[width=1.8cm]{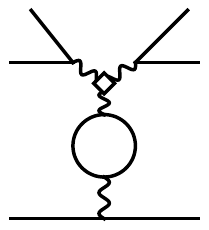}} \quad
\subfigure[]{
	\includegraphics[width=1.5cm]{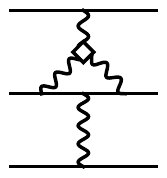}} \quad
\subfigure[]{
	\includegraphics[width=1.8cm]{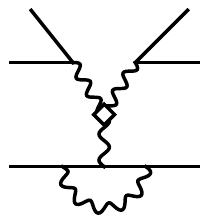}} \quad
\subfigure[]{
	\includegraphics[width=2cm]{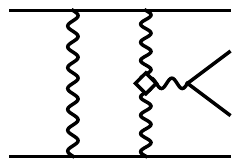}} \quad
\subfigure[]{
	\includegraphics[width=2cm]{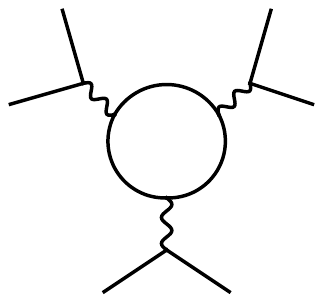}} \quad
\subfigure[]{
	\includegraphics[width=2cm]{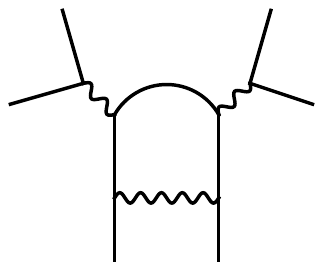}} \quad
\subfigure[]{
	\includegraphics[width=1.8cm]{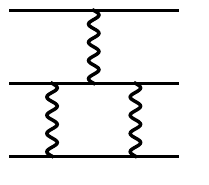}} \quad
\subfigure[]{
	\includegraphics[width=2cm]{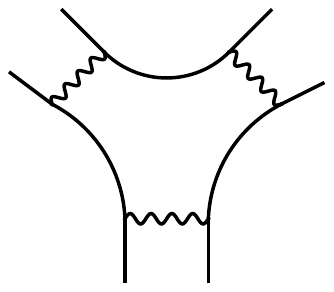}} 
\caption{\label{6P}Feynman diagrams contribute to the beta equation of the six-point vertex.}
\end{figure}

Since there are two vertices, the structure of Feynman diagrams is
\bea
	L &=& I-V_4-V_6 + 1, \\
	E &=& 4 V_4 + 6V_6  - 2I,
\eea
where $L, I$, and $E$ denote the number of loops, internal propagators, and eternal propagators, and $V_4$ and $V_6$ denote the number of four-vertices and six-vertices, respectively. For the RG equation at the one-loop level, the Feynman diagrams contributing to the RG of polarization $(V_4,V_6)=(1,0)$, four-point vertex $(2,0), (0,1)$, and six-point vertex $(1,1), (3,0)$ are shown in Figs.~\ref{2P},~\ref{4P},and~\ref{6P}.

The renormalized action coming from the Feynman diagrams is
\bea
	\delta S &=& \int d^dx \Big[g_4 (2N+4) \int_k G(k) \phi^2 \nn\\
	&&+ g_6(3N+12) \int_k G(k) \phi^4 \nn\\
	&& - \frac12 g_4^2 (8N+64) \int_k G(k)^2 \phi^4 \nn\\
	&&- g_4 g_6 (12N+168) \int_k G(k)^2 \phi^6  \nn\\
	&& + \frac16 g_4^3 (64N + 1664) \int_k G(k)^3 \phi^6 \Big].
\eea
Using the momentum shell, the loop integrals are
\bea
	\int_k G(k) &=& \int_k^\Lambda \frac1{k^2 + r} = \frac{A_{d-1}}{(2\pi)^d} \frac{\Lambda^2l}{1+ \bar r}, \\
	\int_k G(k)^2 &=& \int_k^\Lambda \frac1{(k^2 + r)^2} = \frac{A_{d-1}}{(2\pi)^d} \frac{l}{(1+ \bar r)^2}, \\
	\int_k G(k)^3 &=& \int_k^\Lambda \frac1{(k^2 + r)^3} = \frac{A_{d-1}}{(2\pi)^d} \frac{l}{\Lambda^2(1+ \bar r)^3},
\eea
where $\int_k^\Lambda \equiv \int_\Lambda \frac{d^dk}{(2\pi)^d} = \frac{A_{d-1}}{(2\pi)^d} \int_{\Lambda e^{-l}}^\Lambda k^{d-1}dk$, $l>0$ is the running parameter, and $\bar r= r \Lambda^{-2} $. $A_d$ is the area of the $d$-dimensional unit sphere. Using these integrals, we arrive at the effective action
\bea
	 S &=& \int d^dx \Big[ \frac12 \partial \phi^2 + \big( \frac12 r + g_4 (2N+4) \frac{A_{d-1}}{(2\pi)^d} \frac{\Lambda^2l}{1+ \bar r} \big) \phi^2 \nn \\
	 && + \Big( g_4+ g_6(3N+12) \frac{A_{d-1}}{(2\pi)^d} \frac{\Lambda^2l}{1+ \bar r} \nn\\
	 && - \frac12 g_4^2 (8N+64) \frac{A_{d-1}}{(2\pi)^d} \frac{l}{(1+ \bar r)^2} \Big) \phi^4 \nn \\
	 && + \big( g_6 - g_4 g_6 (12N+168) \frac{A_{d-1}}{(2\pi)^d} \frac{l}{(1+ \bar r)^2}  \nn \\
	 && + \frac16 g_4^3 (64N + 1664) \frac{A_{d-1}}{(2\pi)^d} \frac{l}{\Lambda^2(1+ \bar r)^3}  \big) \phi^6\Big],
\eea
which leads to the beta equations in Eqs.~(\ref{RG2a}) and~(\ref{RG2b}) by setting $\bar r = 0$ to restrict to the critical surface.

\section{Zero-dimensional $\phi^4+\phi^6$ theory: a toy model}\label{app:0D}
Here, we study a toy model~\cite{Zinn-Justin1981} to better illustrate the field theory calculation. 
The following integral is a zero-dimension `toy' version of the $\phi^4$ theory:
\bea\label{Z4}
	Z(g_4) = \frac1{\sqrt{2\pi}} \int dx e^{-(\frac12 x^2+ \frac{g_4}4 x^4)}.
\eea
The integral can be done analytically, 
\bea
	Z(g_4) = \frac{e^{\frac{1}{8g_4}}K_{\frac14}(\frac{1}{8g_4})}{2\sqrt{\pi g_4}},
\eea
where $K$ is the modified Bessel function of the second kind. It is apparent that $Z(g_4)$ has a branch cut at $g_4\in (-\infty, 0)$ if one analytically continues $g_4$ to the complex plane. As a result, the integral is controlled by the branch cut, namely,
\bea
	Z(g_4) = \int_{-\infty}^0 \frac{dz}{2\pi i} \frac{ Z(z + i \epsilon) - Z(z - i\epsilon)}{z - g_4} = \int_{-\infty}^0 \frac{dz}{\pi} \frac{\Im Z(z) }{z - g_4}. \nn \\
\eea

We are interested in getting the asymptotic expansion at the strong-coupling limit,
\bea
	Z(g_4) = \sum_{k=0}^\infty A_k g_4^k, \quad k\rightarrow \infty.
\eea
To get the answer, it is straightforward to expand the integral
\bea
	Z(g_4) = \frac1{\sqrt{2\pi}} \int dx e^{-\frac12 x^2} \sum_k \frac{(-1)^k}{k!}4^{-k} x^{4k} g_4^k,
\eea
from which
\bea\label{c4}
	A_k 	= \frac{(-1)^k}{\sqrt\pi} \frac{\Gamma(2k+ \frac12)}{k!} \rightarrow \frac{(-1)^k}{\sqrt2 \pi} 4^k (k-1)! ,
\eea
where we have used the Stirling's formula and the identity
\bea
	\Gamma(n+\frac12) = \frac{(2n)!}{4^n n!} \sqrt \pi.
\eea
We see that the asymptotic expansion is not convergent.

There is another way to get the asymptotic expansion, which relates the expansion coefficient to the branch cut of $g_4$. To get the expansion coefficient, we use 
\bea\label{coefficient}
	A_k &=& \Big[ \frac1{k!} \frac{\partial^k}{\partial g_4^k} \int_{-\infty}^0 \frac{dz}{\pi} \frac{\Im Z(z) }{z - g_4} \Big]_{g_4=0} = \int_{-\infty}^0 \frac{dz}{\pi} \frac{\Im Z(z) }{z^{k+1}}, \nn\\
\eea
from which we know the expansion coefficient through the imaginary part of $Z(g_4)$ at the branch cut.

When $g_4$ is negative, one needs to continue the variable $x$ to make the integral converged. For $g_{4\pm} \equiv g_4 \pm \epsilon$, making $x \equiv \rho e^{i\theta}$, the $\phi^4$ potential in Eq.~(\ref{Z4}) is given by
\bea
	\frac{g_{4\pm}}4 x^4 &=& \frac{|g_4|}4 \rho^4 e^{\pm i\pi + i4\theta} \nn\\
	&=& \frac{|g_4|}4 \rho^4 [\cos(\pm \pi + 4\theta) + i\sin(\pm \pi + 4\theta) ],
\eea
Thus, we can use a different contour at $|x| \gg 1$ for $g_{4\pm}$:
\bea
	&& g_4 =g_{4+},\quad  - \frac{3\pi}8 < \theta < - \frac\pi8, \\
	&& g_4 =g_{4-}, \quad  \frac{\pi}8 < \theta <  \frac{3\pi}8. 
\eea

Instead of evaluating the integral directly, we can use the saddle point approximation, which can be extended to the field theory. The saddle point equation is
\bea
	x+g_4 x^3 = 0,
\eea
where the saddle points are $x=0$ and $x= \pm \frac1{\sqrt{-g_4}}$. Apparently, $x=0$ does not contribute to the imaginary part. And the other two saddle points lead to the contribution
\bea
	\Im Z(z) &=& \frac1{2i} \sum_{x_s=\pm \frac1{\sqrt{-z}}} \frac1{\sqrt{2\pi}}  e^{-(\frac12 x_s^2+ \frac{z}4 x_s^4)} \int_{i\infty}^{-i\infty} dx e^{x^2} \nn \\
	&=& - \frac1{\sqrt2}e^{\frac1{4z}}.
\eea
Thus we get the coefficient
\bea\label{Ak}
	A_k &=& -\int_{-\infty}^0 \frac{dz}{\sqrt2\pi} \frac{1}{z^{k+1}}e^{\frac1{4z}} = \frac{ (-1)^{k}}{\sqrt2\pi} 4^k \Gamma(k) \nn\\
	&=& \frac{ (-1)^{k}}{\sqrt2\pi} 4^k (k-1)!.
\eea
It is precisely the asymptotic expansion shown in Eq.~(\ref{c4}).
From this lesson, we know that it is the unstable saddle point giving rise to the imaginary contribution, and consequently to the asymptotic expansion.

Now let us consider another integral that is the zero-dimensional version of the $\phi^4+\phi^6$ theory,
\bea
	Z(g_4,g_6) = \frac1{\sqrt{2\pi}} \int dx e^{-S(x)}, \\
	S(x) = \frac12 x^2+ \frac{g_4}4 x^4 + \frac{g_6}{6} x^6.
\eea
Unlike the zero-dimensional $\phi^4$ integral, $Z(g_4,g_6)$ is analytic for all $g_4$ as long as $g_6>0$. The branch cut comes from $g_6 \in (-\infty, 0)$. So we can use the same strategy to get the asymptotic expansion coefficient of $g_6$. The asymptotic expansion of $Z(g_4, g_6)$ is given by
\bea\label{expansion}
	Z(g_4, g_6) \rightarrow  \sum_{k} A_k g_4^k + \sum_{k'>0} B_{k'}(g_4) g_6^{k'}, 
\eea
where $A_k$ is given by Eq.~(\ref{Ak}). Similar to Eq.~(\ref{coefficient}), the asymptotic expansion of $g_6$ is given by
\bea
	B_k(g_4) =   \int_{-\infty}^0 \frac{dz}{\pi} \frac{\Im Z(g_4, z) }{z^{k+1}}.
\eea
We can use the saddle-point approximation to get the imaginary part of the integral. 
In the following, we consider $g_4$ as a small perturbation. The saddle-point equation is ($g_4=0, g_6<0$)
\bea
	x + g_6 x^5 =0,
\eea
and the saddle-points are (for $g_6<0$) $x=0, x_c =\pm \frac1{(-g_6)^{1/4}}$. 
The quadratic fluctuation near the saddle point solution is given by
\bea
	S(\delta x) & \approx &  -\frac{g_4}{4g_6} + \frac1{3\sqrt{-g_6}} -2\delta x^2,
\eea
where we include $\frac{g_4}4 x^4$ as a perturbation. 
As expected from the unstable solution, the fluctuation has a negative mass. This contributes to the imaginary part of integral, namely, the imaginary part of the integral is given by
\bea
	\Im Z(g_4, g_6) &=& \frac1{2i} \sum_{x=x_c}\frac1{\sqrt{2\pi}} \int_{i\infty}^{-i \infty} d\delta x e^{-S(\delta x)}  \nn\\ 
	&=& -\frac12 e^{\frac{g_4}{4g_6} - \frac1{3\sqrt{-g_6}}},
\eea
from which we can get the asymptotic expansion 
\begin{eqnarray}\label{Bk}
	B_k(g_4) =\frac{(-1)^k}{\pi} g_4^{-k} \Gamma(2k)U \Big(k, \frac12, \frac1{9g_4} \Big),
\end{eqnarray}
where $U$ is the confluent hypergeometric function.

Thus the theory has at least two fates. On the one hand, if $g_4 \gg g_6$, the asymptotic expansion coefficient of $g_6$ vanishes,
\bea
	\lim_{g_4 \rightarrow \infty} B_k(g_4)=0.
\eea
The expansion is controlled by $A_k$ in Eq.~(\ref{Ak}) as expected. On the other hand, if $g_4 \ll g_6$, we recover the asymptotic expansion of the pure $\phi^6$ integral,
\bea
	\lim_{g_4 \rightarrow 0} B_k(g_4) = \frac{(-1)^k}\pi 9^k \Gamma(2k) \rightarrow \frac{(-1)^k 6^{2k}}{2(\pi k)^{3/2}} (k!)^2.
\eea
To check this, it is straightforward to calculate the coefficient directly,
\bea
	B_k(0) &=& \frac{(-\frac16)^k}{k!} \frac1{\sqrt{2\pi}}\int_{-\infty}^\infty dx x^{6k} e^{-\frac12 x^2} \nn\\
	&=& \frac{(- \frac43)^k}{\sqrt\pi} \frac{\Gamma(3k+\frac12)}{\Gamma(k+1)} \rightarrow \frac{(-1)^k 6^{2k}}{2(\pi k)^{3/2}} (k!)^2.
\eea
We can see that the results from the saddle-point approximation exactly reproduce the asymptotic expansion. More importantly, different from the $\phi^4$ integral, the asymptotic coefficient of the $\phi^6$ integral increases as $(k!)^2$, which is qualitatively different from that of the $\phi^4$ theory.

\section{Saddle-point solutions of $\phi^4$ and $
\phi^6$ theories}\label{app:saddlepoint}

The saddle-point equation for pure, massless $\phi^4$ theory is
\begin{equation}
    \partial^2\xi+\xi^3=0.\label{phi4saddles}
\end{equation}
We assume that $\xi(x)$ is spherically symmetric, i.e., it only depends on $x^2=\sum_{i=1}^dx_i^2$, where $d$ is the dimension of the (Euclidean) spacetime. Writing $f(y)=\xi(x)$ where $y\equiv x^2$, Eq.~(\ref{phi4saddles}) becomes
\begin{equation}
    4yf''(y)+2df'(y)+f^3(y)=0.\label{feqn}
\end{equation}
In $d=4$ dimensions, the solution to this equation is
\begin{equation}
    f(y)=\sqrt{\frac{c+2}{y}}\hbox{sn}\left(\frac{\sqrt{c}}{2\sqrt{2}}\log(y/y_0)\bigg|\frac{c+2}{c}\right),\label{jacobi}
\end{equation}
where $\hbox{sn}(w,m)$ is the Jacobi elliptic function, and $y_0$ and $c$ are integration constants. For $c\approx0$, this reduces to
\begin{equation}
    f(y)=\frac{16\sqrt{cy_0}}{cy+32y_0}.
\end{equation}
If we then set $c=\pm2$ and $y_0=\pm1/16$, this becomes
\begin{equation}
    f(y)=\pm\frac{2\sqrt{2}}{1+y}.\label{canonicalsoln}
\end{equation}
The fact that we can set $c=\pm2$ in the last step and still obtain a valid solution is a consequence of the scale invariance of Eq.~(\ref{phi4saddles}). Does Eq.~(\ref{jacobi}) contain other valid solutions beyond Eq.~(\ref{canonicalsoln})? To answer this question, we return to Eq.~(\ref{feqn}) and analyze the large $y$ behavior. As $y\to\infty$, we require $f(y)\to0$ because we need the action at the saddle point, $S_0=\int d^dx\xi^4(x)$, to be finite so that it makes a non-negligible contribution to the correlation function. In this limit, the $f^3(y)$ term either decays more quickly than the other two terms in Eq.~(\ref{feqn}) or it is comparable to them. If it is comparable, then the solution behaves as 
\begin{equation}
    f(y)\to \pm\frac{\sqrt{d-3}}{\sqrt{y}}\quad\hbox{as}\quad y\to\infty.
\end{equation}
This implies that $\xi(x)\sim1/|x|$ as $|x|\to\infty$, which in turn leads to a divergent action, $S_0\to\infty$, for dimension $d\ge4$. Therefore, we discard such solutions and instead consider the case where the $f^3(y)$ decays faster than the other terms in Eq.~(\ref{feqn}). In this case, the asymptotic behavior of $f(y)$ is
\begin{equation}
    f(y)\to cy^{1-d/2}\quad\hbox{as}\quad y\to\infty,\label{fasymptotic}
\end{equation}
where $c$ is a constant. This yields $\xi(x)\sim|x|^{2-d}$ as $|x|\to\infty$, and the action is finite for $d\ge3$. When $d=4$, this asymptotic behavior is of course the same as that of Eq.~(\ref{canonicalsoln}). The constant $c$ in Eq.~(\ref{fasymptotic}) is again a reflection of the scale-invariance of the solutions, and so we conclude that the only two solutions with finite action are those given in Eq.~(\ref{canonicalsoln}).

The saddle-point equation for pure, massless $\phi^6$ theory is
\begin{equation}
    \partial^2\chi+\chi^5=0.\label{phi6saddle}
\end{equation}
We again assume that $\chi(x)$ is spherically symmetric, in which case Eq.~(\ref{phi6saddle}) reduces to
\begin{equation}
    4yg''(y)+2dg'(y)+g^5(y)=0,\label{geqn}
\end{equation}
where $g(y)=\chi(x)$. We require that the solutions vanish as $y\to\infty$ so that $S_6=\int d^dx\chi^6(x)<\infty$. If $g^5(y)$ is comparable to the other two terms in Eq.~(\ref{geqn}) as $y\to\infty$, then the asymptotic behavior is
\begin{equation}
    g(y)\to\pm\frac{(2d-5)^{1/4}}{\sqrt{2}y^{1/4}}.
\end{equation}
This implies that $\chi(x)\sim1/\sqrt{|x|}$ as $|x|\to\infty$, which in turn leads to a divergent $S_6$ for $d\ge3$. We therefore conclude that the $g^5(y)$ term decays faster than the other two terms in Eq.~(\ref{geqn}), which then yields the following asymptotic behavior:
\begin{equation}
    g(y)\to cy^{1-d/2}\quad\hbox{as}\quad y\to\infty.
\end{equation}
Because Eq.~(\ref{phi6saddle}) has a scale invariance such that if $\chi(x)$ is a solution then so is $b^{-1}\chi(x/b^2)$, we understand that $c$ is a redundant scale factor. Therefore, there are only two distinct solutions of Eq.~(\ref{phi6saddle}) that contribute to correlation functions, and these two solutions are related by an overall minus sign.

\section{Quasi-local approximation of a determinant}\label{app:det}
First consider a determinant~\cite{Zinn-Justin1981}
\bea
	\det(-\tri + m^2 + g V(x)) (-\tri + m^2)^{-1}.
\eea
where $\tri = \sum_\mu^d \partial_\mu^2$. 
A useful formula is
\bea
	\Tr \log A B^{-1} = - \Tr \int_0^\infty \frac{dt}t (e^{-At} - e^{-Bt}).
\eea
Since we are interested in the large $g$ behavior, the integral is dominated by the small $t$ regime. We can use the quasilocal~\cite{Parisi1977a, Zinn-Justin1981} approximation,
\bea
	e^{-(-\tri + m^2 + g V(x)) t} \approx e^{-t(-\tri + m^2)} e^{-t g V(x)}.
\eea
Now we have
\bea
	&& \Tr \log (-\tri + m^2 + g V(x)) (-\tri + m^2)^{-1} \nn \\
	&=& - \Tr \int_0^\infty \frac{dt}t e^{-t(-\tri + m^2)} (e^{-t g V(x)}- 1) \nn \\
	&=& - \int_0^\infty \frac{dt}t \int \frac{d^dp}{(2\pi)^d}e^{-t(p^2 + m^2)} \int d^d x (e^{-t g V(x)}- 1)  \nn\\
	&=& - \frac1{(4\pi)^{d/2}} \int d^d x \int_0^\infty \frac{dt}t \frac1{t^{d/2}} [e^{-t (m^2+ g V(x))} -e^{-tm^2}]   \nn\\
\label{det}	&=& - \frac{\Gamma(-d/2)}{(4\pi)^{d/2}} \int d^d x  [(m^2+ g V(x))^{d/2} - m^d]. 
\eea
One can also evaluate the correction to the above approximation order by order using the Baker–Campbell–Hausdorff formula. 
But we do not consider these corrections here.

Now we include spin, and consider the determinant of the Dirac operator,
\bea
	\mathcal D(e) = \det( \slashed \partial + i e \phi \gamma^0 + m)( \slashed \partial +  m)^{-1} .
\eea
where $\slashed \partial = \gamma^\mu \partial_\mu$, and $\gamma^\mu$ is the $4\times4$ Dirac matrix. 
Here to regularize the determinant, we introduce a mass $m$. 
We set it to zero in the final step to retain the gapless Dirac dispersion. 
To evaluate the determinant, note that under the parity symmetry,
\bea
	\mathcal D(e) \rightarrow \det( -\slashed \partial - i e \phi \gamma^0 + m)( -\slashed \partial +  m)^{-1} .
\eea
As a result, we can evaluate the square of determinant, i.e., 
\bea
	\log \mathcal D(e)^2 = \Tr \log ( -\tri + e^2 \phi^2 + m^2)( -\tri +  m^2)^{-1}, \nn\\
\eea
where we have assumed the field $\phi$ is large and smooth to neglect the correction from derivatives and used the fact the $\gamma$ matrix is traceless.
From Eq.~(\ref{det}), we note 
\bea
	\mathcal D(e) = \exp \Big[- \frac{2\Gamma(-d/2)}{(4\pi)^{d/2}} \int d^d x  [(m^2+ e^2 \phi^2)^{d/2} - m^d] \Big]. \nn\\
\eea
Now we can safely send $m$ to zero to get the effective field theory in Eq.~(\ref{detEff}). Including the Fermi velocity and the fermion species $N$, the final answer is modified as
\bea
	\mathcal D(e) = \exp \Big[- \frac{2\Gamma(-d/2)}{(4\pi)^{d/2}} \frac{N}{v^{d-1}} \int d^d x  [(m^2+ e^2 \phi^2)^{d/2} - m^d] \Big]. \nn\\
\eea


\begin{thebibliography}{99}

\bibitem{Nio2018} T. Aoyama, T. Kinoshita, and M. Nio, {\it Revised and improved value of the QED tenth-order electron anomalous magnetic moment}, Phys. Rev. D {\bf97}, 036001 (2018).

\bibitem{Schwinger1948} J. Schwinger, {\it On Quantum-Electrodynamics and the Magnetic Moment of the Electron}, Phys. Rev. {\bf73}, 416 (1948).

\bibitem{Landau1954} L. D. Landau, A. A. Abrikosov, and I. M. Khalatnikov, {\it On the removal of infinities in quantum electrodynamics}, Dokl. Akad. Nauk SSSR {\bf95}, 497 (1954).

\bibitem{LandauPole} See \url{https://en.wikipedia.org/wiki/Landau_pole}.

\bibitem{Stuben1998} M. G\"ockeler, R. Horsley, V. Linke, P. Rakow, G. Schierholz, and H. St\"uben, {\it Is There a Landau Pole Problem in QED?}, Phys. Rev. Lett. {\bf80}, 4119 (1998).

\bibitem{Gies2004} H. Gies and J. Jaeckel, {\it Renormalization Flow of QED}, Phys. Rev. Lett. {\bf93}, 110405 (2004).

\bibitem{Djukanovic2018} D. Djukanovic, J. Gegelia, and Ulf-G. Mei\ss ner, {\it Triviality of quantum electrodynamics revisited}, Commun. Theor. Phys. {\bf 69} (2018) 263. 

\bibitem{Dashen1983} R. Dashen and H. Neuberger, {\it How to Get an Upper Bound on the Higgs Mass}, Phys. Rev. Lett. {\bf50}, 1897 (1983).

\bibitem{Weisz} M. L\"uscher and P. Weisz, {\it Scaling laws and triviality bounds in the lattice $\phi4$ theory: (I). One-component model in the symmetric phase}, Nucl. Physics. {\bf B290} (1987) 25; {\it Scaling laws and triviality bounds in the lattice $\phi4$ theory: (II). One-component model in the phase with spontaneous symmetry breaking}, Nucl. Physics. {\bf B295} (1988) 65.

\bibitem{Lee1977} B. W. Lee, C. Quigg, and H. B. Thacker,  {\it Strength of Weak Interactions at Very High Energies and the Higgs Boson Mass}, Phys. Rev. Lett. {\bf38}, 883 (1977).

\bibitem{Dunne2015} G. V. Dunne and M. \"Unsal {\it What is QFT? Resurgent trans-series, Lefschetz thimbles, and new exact saddles}, PoS LATTICE 2015 (2016) 010.




\bibitem{Landau1955} L. D. Landau, in W. Pauli, ed. {\it Niels Bohr and the Development of Physics}, (London: Pergamon Press 1955).

\bibitem{Shirkov1976}
N. N. Bogoliubov and D. V. Shirkov, {\it Introduction to the Theory of Quantized Fields}, 3rd ed. (Nauka, Moscow, 1976; Wiley, New York, 1980).




\bibitem{Callaway1988} D. J. E. Callaway, {\it Triviality Pursuit: Can Elementary Scalar Particles Exist?}, Physics Reports {\bf167}, 241 (1988). 
\bibitem{Callaway1986} D. J. E. Callaway and R. Petronzio, {\it Can elementary scalar particles exist?: (II). Scalar electrodynamics}, Nuclear Physics {\bf B277}, 50 (1986). 
\bibitem{Kim2002} S. Kim, J. B. Kogut, and  M.-P. Lombardo, {\it Gauged Nambu–Jona-Lasinio studies of the triviality of quantum electrodynamics}, Phys. Rev. D {\bf65}, 054015 (2002). 


\bibitem{Lizzi2013} A. A. Andrianov, D. Espriu, M. A. Kurkov, and F. Lizzi, {\it Universal Landau Pole}, Phys. Rev. Lett. {\bf111}, 011601 (2013).

 
\bibitem{Lipatov1977} L. N. Lipatov, {\it Divergence of the perturbation-theory series and the quasiclassical theory}, Zh. Eksp. Teor. Fi. {\bf72}, 411 (1977).
\bibitem{Zinn-Justin1981} J. Zinn-Justin, {\it Perturbation series at large orders in quantum mechanics and field theories: Application to the problem of resummation}, Physics Reports {\bf70}, 109 (1981).
\bibitem{Suslov2001} I. M. Suslov, {\it Summing divergent perturbative series in a strong coupling limit. The Gell-Mann-Low function of the $\phi^4$ theory}, Journal of Experimental and Theoretical Physics, {\bf93}, 1 (2001).
\bibitem{Suslov2008} I. M. Suslov, {\it Renormalization group functions of the $\phi^4$ theory in the strong coupling limit: Analytical results}, Journal of Experimental and Theoretical Physics {\bf107}, 413 (2008). 

\bibitem{Kadanoff1966} L. P. Kadanoff, {\it Scaling laws for Ising models near $T_c$}, Physics {\bf 2}, 263 (Long Island City, N.Y. 1966). 
\bibitem{Wilson1975} K. G. Wilson, {\it The renormalization group: critical phenomena and the Kondo problem}, Rev. Mod. Phys. {\bf47}, 773 (1975). 
\bibitem{Wilson1971} K. G. Wilson, {\it Renormalization group and critical phenomena. I. Renormalization group and the Kadanoff scaling picture}, Phys. Rev. B {\bf4}, 3174 (1971).

\bibitem{Geim2009} A. H. Castro Neto, F. Guinea, N. M. R. Peres, K. S. Novoselov, and A. K. Geim, {\it The electronic properties of graphene}, Rev. Mod. Phys. {\bf81}, 109 (2009).

\bibitem{MacDonald2011} R. Bistritzer, and A. H. MacDonald, {\it Moir\'e bands in twisted double-layer graphene}, Proceedings of the National Academy of Sciences {\bf108}, 12233 (2011). 

\bibitem{Cao2018a} Y. Cao, V. Fatemi, S. Fang, K. Watanabe, T. Taniguchi, E. Kaxiras, and P. Jarillo-Herrero, {\it Unconventional superconductivity in magic-angle graphene superlattices}, Nature {\bf556}, 43 (2018).

\bibitem{Cao2018b} Y. Cao, V. Fatemi, A. Demir, S. Fang, S. L. Tomarken, J. Y. Luo, J. D. Sanchez-Yamagishi, K. Watanabe, T. Taniguchi, E. Kaxiras, R. C. Ashoori, and P. Jarillo-Herrero, {\it Correlated insulator behaviour at half-filling in magic-angle graphene superlattices}, Nature {\bf556}, 80 (2018).

\bibitem{SDS2019} S. Das Sarma, and F. Wu, {\it Electron-phonon and electron-electron interaction effects in twisted bilayer graphene}, arXiv:1910.08556.


\bibitem{Scammell2015} H. D. Scammell and O. P. Sushkov, {\it Asymptotic freedom in quantum magnets}, Phys. Rev. B {\bf92}, 220401(R) (2015).

\bibitem{Vozmediano1994} J. Gonz\'alez, F. Guineai, and M. A. H. Vozmediano, {\it Non-Fermi liquid behavior of electrons in the half-filled honeycomb lattice (A renormalization group approach)}, Nucl. Phys. {\bf B424}, 595 (1994).
\bibitem{Vozmediano2011} M. A. H. Vozmediano, {\it Graphene: The running of the constants}, Nat. Phys. {\bf7}, 671 (2011).

\bibitem{Elias2011} D. C. Elias, R. V. Gorbachev, A. S. Mayorov, S. V. Morozov, A. A. Zhukov, P. Blake, L. A. Ponomarenko, I. V. Grigorieva, K. S. Novoselov, F. Guinea, and A. K. Geim, {\it Dirac cones reshaped by interaction effects in suspended graphene}, Nat. Phys. {\bf7}, 701 (2011).


\bibitem{Son2007} D. T. Son, {\it Quantum critical point in graphene approached in the limit of infinitely strong Coulomb interaction}, Phys. Rev. B {\bf75}, 235423 (2007).

\bibitem{SDS2014} E. Barnes, E. H. Hwang, R. E. Throckmorton, and S. Das Sarma, {\it Effective field theory, three-loop perturbative expansion, and their experimental implications in graphene many-body effects}, Phys. Rev. B {\bf89}, 235431 (2014).

\bibitem{Borel1928} E. Borel, {\it Leccecdon sur les series divergentes}, (Gauthier Villars, Paris 1928).

\bibitem{Hofmann2014} J. Hofmann, E. Barnes, and S. Das Sarma, {\it Why Does Graphene Behave as a Weakly Interacting System?}, Phys. Rev. Lett. {\bf113}, 105502 (2014).

\bibitem{Parisi1977a} G. Parisi, {\it Asymptotic Estimates in Perturbation Theory with Fermions}, Phys. Lett. {\bf66B}, 382 (1977).

\bibitem{Parisi1977b} C. Itzykson, G. Parisi, and J. B. Zuber, {\it Asymptotic estimates in quantum electrodynamics}, Phys. Rev. D {\bf16}, 996 (1977).
\bibitem{Parisi1978} R. Balian, C. Itzykson, G. Parisi, and J. B. Zuber, {\it Asymptotic estimates in quantum electrodynamics. II}, Phys. Rev. D {\bf17}, 1041 (1978).


\bibitem{Sigrist2002} M. Matsumoto, B. Normand, T. M. Rice, and M. Sigrist, {\it Magnon Dispersion in the Field-Induced Magnetically Ordered Phase of TlCuCl3}, Phys. Rev. Lett. {\bf89}, 077203 (2002).
\bibitem{Boehm2008} Ch. R\"uegg, B. Normand, M. Matsumoto, A. Furrer, D. F. McMorrow, K. W. Kr\"amer, H. -U. G\"udel, S. N. Gvasaliya, H. Mutka, and M. Boehm, {\it Quantum Magnets under Pressure: Controlling Elementary Excitations in TlCuCl3}, Phys. Rev. Lett. {\bf100}, 205701 (2008).
\bibitem{Sushkov2011} Y. Kulik and O. P. Sushkov, {\it Width of the longitudinal magnon in the vicinity of the O(3) quantum critical point}, Phys. Rev. B {\bf84}, 134418 (2011).

\bibitem{Young2019} H. Polshyn, M. Yankowitz, S. Chen, Y. Zhang, K. Watanabe, T. Taniguchi, C. R. Dean, and A. F. Young, {\it Large linear-in-temperature resistivity in twisted bilayer graphene}, Nature Physics {\bf 15}, 1011 (2019).

\end{thebibliography}
\end{document}